\documentclass[sigconf]{acmart}

\AtBeginDocument{%
  \providecommand\BibTeX{{%
    \normalfont B\kern-0.5em{\scshape i\kern-0.25em b}\kern-0.8em\TeX}}}

\copyrightyear{2020} 
\acmYear{2020} 
\setcopyright{acmcopyright}\acmConference[CCSW'20]{2020 Cloud Computing Security Workshop}{November 9, 2020}{Virtual Event, USA}
\acmBooktitle{2020 Cloud Computing Security Workshop (CCSW'20), November 9, 2020, Virtual Event, USA}
\acmPrice{15.00}
\acmDOI{10.1145/3411495.3421352}
\acmISBN{978-1-4503-8084-3/20/11}

\usepackage{tikz}

\usepackage{hyperref}
\hypersetup{
   colorlinks=true,
   citecolor=teal
}

\usepackage{subfigure}
\usepackage{graphicx}
\usepackage{enumitem}
\usepackage{array}
\usepackage{amsmath, amsfonts, amssymb, amsthm}
\usepackage{xspace}     
\usepackage{booktabs}     
\usepackage{color, colortbl}

\usepackage{algorithm2e}
\newtheorem{remark}{Remark}
\theoremstyle{definition}

\theoremstyle{definition}

\newenvironment {squishlist}
{\begin{list}{$\bullet$}
  { \setlength{\itemsep}{1pt}
     \setlength{\parsep}{1pt}
     \setlength{\topsep}{1pt}
     \setlength{\partopsep}{1pt}
     \setlength{\leftmargin}{1.5em}
     \setlength{\labelwidth}{1em}
     \setlength{\labelsep}{0.5em} } }
{\end{list}}

\renewcommand{\theenumii}{\theenumi.\arabic{enumii}.}

\newcommand{\DP}{\emph{DP}\xspace}

\newcommand{\ML}{ML\xspace}

\newcommand{\LR}{\emph{LR}\xspace}
\newcommand{\NB}{\emph{NB}\xspace}

\newcommand{\NN}{\emph{NN}\xspace}

\newcommand{\ACL}{\emph{ACL}\xspace}
\newcommand{\ShokriMI}{\emph{ShokriMI}\xspace}
\newcommand{\SalemMI}{\emph{SalemMI}\xspace}
\newcommand{\SalemAI}{\emph{SalemAI}\xspace}
\newcommand{\YeomMI}{\emph{YeomMI}\xspace}
\newcommand{\YeomAI}{\emph{YeomAI}\xspace}



\begin{document}

\date{}

\title{Not one but many Tradeoffs: Privacy Vs. Utility in Differentially Private Machine Learning}

\author{Benjamin Zi Hao Zhao}
\email{benjamin.zhao@unsw.edu.au}
\affiliation{%
  \institution{University of New South Wales and Data61 CSIRO}
  \city{Sydney}
  \country{Australia}
}
\author{Mohamed Ali Kaafar}
\email{dali.kaafar@mq.edu.au}
\affiliation{%
  \institution{Macquarie University}
  \city{Sydney}
  \country{Australia}
}
\author{Nicolas Kourtellis}
\email{nicolas.kourtellis@telefonica.com}
\affiliation{%
  \institution{Telefonica Research}
  \city{Barcelona}
  \country{Spain}
}

\setlength{\textfloatsep}{5pt}
\settopmatter{printfolios=true}
\pagenumbering{arabic}

\begin{abstract}


Data holders are increasingly seeking to protect their user's privacy, whilst still maximizing their ability to produce machine learning (ML) models with high quality predictions.
In this work, we empirically evaluate various implementations of differential privacy (DP), and measure their ability to fend off real-world privacy attacks, in addition to measuring their core goal of providing accurate classifications.
We establish an evaluation framework to ensure each of these implementations are fairly evaluated.
Our selection of DP implementations add DP noise at different positions within the framework, either at the point of data collection/release, during updates while training of the model, or after training by perturbing learned model parameters.
We evaluate each implementation across a range of privacy budgets and datasets, each implementation providing the same mathematical privacy guarantees.
By measuring the models' resistance to real world attacks of membership and attribute inference, and their classification accuracy. we determine which implementations provide the most desirable tradeoff between privacy and utility.
We found that the number of classes of a given dataset is unlikely to influence where the privacy and utility tradeoff occurs, a counter-intuitive inference in contrast to the known relationship of increased privacy vulnerability in datasets with more classes.
Additionally, in the scenario that high privacy constraints are required, perturbing input training data before applying ML modeling does not trade off as much utility, as compared to noise added later in the ML process.


\end{abstract}

\maketitle

\section{INTRODUCTION}\label{sec:introduction}

Advanced machine learning (ML) techniques enable accurate data analytics for various application domains. This promoted the commercial deployment of ML as a service (offered by data giants, such as Google and Amazon) which allows data-driven businesses to train models on sensitive data while offering third party (paid) access to these models. Although commercially attractive, these services can be vulnerable to model theft and privacy infringements potentially not compliant with developing privacy regulations (e.g., EU and USA regulations such as COPPA~\cite{COPPA1998} and GDPR~\cite{GDPR2018}, and most recently e-Privacy~\cite{ePrivacy2019} and CCPA~\cite{CCPA2020}). 
In order to preserve their models' privacy while still maximizing their ability to produce ML and deep learning (DL) models that have high utility for their services, data-driven organizations are turning towards leveraging privacy-preserving ML (PPML) techniques, building on theoretical frameworks of Differential Privacy~\cite{dwork2006our, dwork2014algorithmic} (\DP) and/or Federated Learning~\cite{konecny15fl} (FL). However, differentially private PPML methods often come with an intrinsic tradeoff between utility (e.g., as captured by accuracy of the model) and the privacy guarantees offered by the technique applied to protect user data.

A recent initial investigation in~\cite{jayaraman2019evaluating} studies different \DP compositions, and how these compositions can be applied to the training of a neural network or logistic regression model. \cite{jayaraman2019evaluating} reports on the impact these privacy mechanism have on the model's utility, and the effectiveness of inference attacks on the resulting models.
Inspired by~\cite{jayaraman2019evaluating}, and towards the goal of understanding the tradeoff between privacy and utility of \DP-enabled ML methods, we dive deeper into this problem and, in this study, we set to assess how this inherent tradeoff depends on the (1) ML method used, (2) stage in the ML framework where the \DP method is applied to protect the data or model, and (3) complexity of training data in use with respect to classes and attributes in the data.

We develop a comprehensive and systematic evaluation of a \DP-enabled ML framework that enables a privacy ML researcher to study the Utility-Privacy tradeoff in depth for their data at hand. Our objective is to allow the selection of the best performing method yielding the highest predictive accuracy while still ensuring a solid level of privacy protection, by studying the different stages where \DP-based noise can be applied: as an obfuscation to the input data, during model training, or at the model finalization by perturbing the learned model parameters. Equally important, the study's objective is to inform privacy ML researchers what privacy threshold to apply in their framework, and what are the privacy guarantees expected from the selected setup, vs. the utility of the chosen ML method.

We study various recent \DP implementations of classical ML and DL methods such as Naive Bayes, and Neural Networks, and empirically measure their ability to fend off black-box privacy attacks that may be practically launched in the real-world, while also measuring the model's core goal of providing accurate classifications.
Crucially, we establish this standard evaluation framework to ensure each of these \DP implementations are evaluated fairly.

In particular, we study and test how ML performance and privacy are impacted when \DP noise is added at different stages of the ML pipeline:
Stage (1) by adding noise to the input data before the ML/DL training phase. 
Stage (2) where \DP noise is added during model updates, i.e., while training the selected model.
Stage (3) after the model training is performed, by perturbing learned model parameters.
We evaluate each \DP-enabled ML implementation across a range of privacy budgets, each instance providing the same mathematical privacy guarantees.
We measure different metrics to capture the aforementioned tradeoff: privacy offered to the model and data (resistance to membership and attribute inference attacks) and model utility (classification accuracy).

We use both synthetic and real-world datasets to capture the aforementioned privacy and utility tradeoff.
Our use of a synthetic dataset enables us to isolate the effects of \DP noise, stages and dataset complexity without the influence of data distributions. However, not to discount the importance of standard real-world datasets, we also perform our evaluation on a range of real data like CIFAR~\cite{krizhevsky2009learning}, Purchase~\cite{purchase_dataset}, and the Netflix dataset~\cite{netflix_dataset} in which we provide the same pre-processing treatment as Purchase~\cite{purchase_dataset}.

With our experimentation, we make the following observations.
Most notably, for a given amount of model utility, applying \DP noise at stages later than the input phase permits the addition of more \DP noise, thus providing higher privacy guarantees. This observation is consistent across all \DP-ML algorithms.

When considering utility and privacy as function of the \DP noise, we identify an ``inflection point'' for each function, an indicator of where the greatest change in utility and/or privacy will occur for a given \DP-ML method.
We find that this point on privacy function is more closely related to the Utility response, and the \DP-ML method used, instead of \DP privacy guarantees, as expected from the amount of \DP noise applied to the process.
%
We demonstrate, though common knowledge, that class complexity has an impact on the absolute performance of a privacy attack. However, counter-intuitively, we observe that this complexity (number of classes) does not influence the inflection point of the utility of the privacy function. 
Finally, when privacy or utility come with constraints, we provide recommendations for best performing \DP-ML method, and their expected utility and privacy guarantees.

We contribute our open sourced framework\footnote{Source code available at: \url{https://github.com/PrivateUtility/PrivateUtility}} for reproducibility purposes, as well as for other researchers to build on it and study privacy and utility thresholds of newly proposed \DP-ML methods.

\section{Methodology}\label{sec:methodology}

\begin{figure*}[t]
	\centering
	\includegraphics[width=0.90\textwidth]{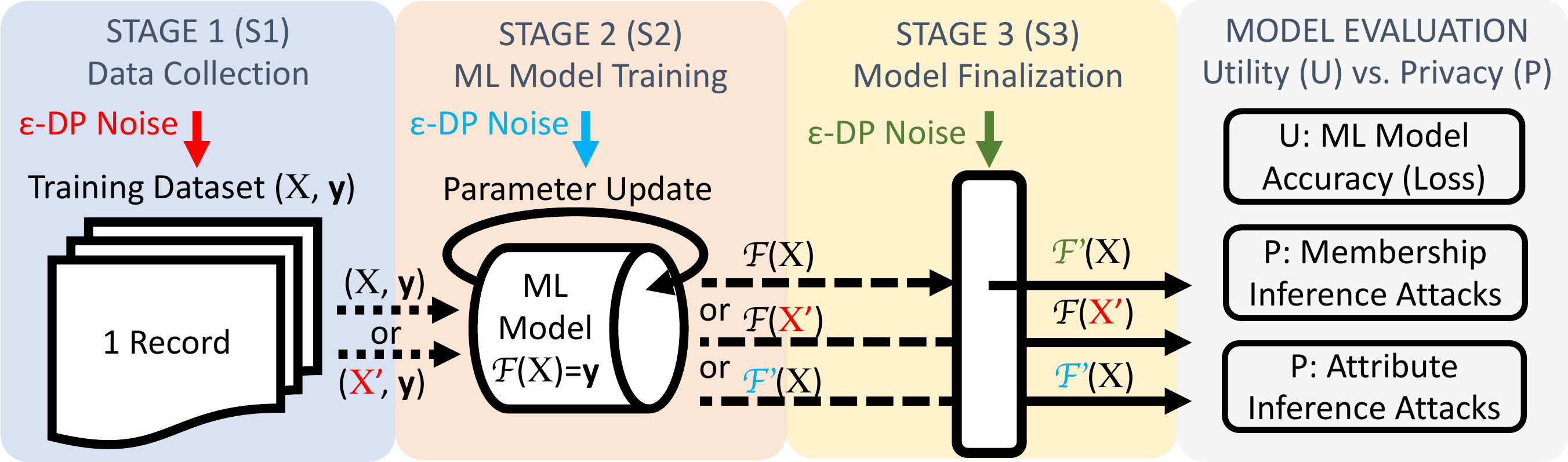}
	\vspace{-4mm}
	\caption{Our instantiation of the proposed methodology, with the three possible Stages that \DP noise can be introduced in the ML pipeline to guarantee data privacy, and performance metrics used to assess privacy-utility tradeoff.}
	\label{fig:eval_pipe}
	\vspace{-2mm}
\end{figure*}

\subsection{Overview}
\label{sec:methodology-overview}

In this Section, we provide details of the building blocks needed to study the privacy-utility tradeoff as a comprehensive and modular methodology. Our methodology encompasses the following:

\begin{squishlist}
\item \DP noise definitions (Sec.~\ref{sec:dp-definitions})
\item Stages of the ML pipeline at which \DP noise is added (Sec.~\ref{sec:stages})
\item ML algorithms studied that are \DP-enabled (Sec.~\ref{sec:algorithms})
\item Privacy metrics, assessed with privacy attacks on data (Sec.~\ref{sec:privacy-attacks})
\item ML utility metrics (Sec.~\ref{sec:utility-metrics})
\end{squishlist}

In this work, we provide an instantiation of this methodology (Figure~\ref{fig:eval_pipe}) to evaluate the privacy-utility tradeoff in \DP-enabled ML algorithms.
Next, we cover details for each of these building blocks, and in Section~\ref{sec:experiments}, we provide details of their implementation. 

Note that our methodology can be extended to account for other considerations in the privacy-utility tradeoff analysis. This could include Resource metrics (e.g., required computational resources for training ML models) or various datasets characteristics in use.

\subsection{Differential Privacy}
\label{sec:dp-definitions}

Differential privacy (\DP) mathematically defines the protection offered in regards to the privacy of a single data vector, whether that is representative of an individual, or a single temporal event~\cite{dwork2006calibrating, dwork2006our}.
The $\epsilon$-differential privacy is defined such that two neighboring sets of data $D$ and $D'$, differing by a single vector are indistinguishable up to a limit as described by a privacy budget $\epsilon$.
The output of a mechanism $\mathcal{M}$ applied on each dataset should also be indistinguishable from each other, up to our limit of $\epsilon$.
In other words:
\begin{equation}
Pr[\mathcal{M}(D) \in S ]  \le Pr[\mathcal{M}(D') \in S ] * e^\epsilon
\end{equation}

\noindent
Many differentially private ML algorithms support relaxations of the \DP definition.
There are two main relevant relaxations of $\epsilon$-\DP: $(\epsilon, \delta)$-\DP~\cite{dwork2014algorithmic}, and $(\alpha, \epsilon)$-\DP (Renyi-DP)~\cite{mironov2017renyi}.
Both relaxations provide eased requirements for \DP, while preserving properties such as composition and core privacy guarantees. 

We will not be using these relaxations in this work, however, we note they are reducible to $(\epsilon)$-\DP, the focus definition in this paper.
In fact,
$(\epsilon, \delta)$-\DP~\cite{dwork2014algorithmic} is equivalent to $(\epsilon)$-\DP, when $\delta = 0$, and 
$(\alpha, \epsilon)$-\DP~\cite{mironov2017renyi} is reduced to $\epsilon$-\DP when $\alpha = \inf$.
Also, the authors in~\cite{bassily2014private}, given a set of assumptions, derive the upper bound of $\epsilon$-differential privacy as $p/\epsilon$, where $p$ is the dataset dimensionality.

\subsection{ML Pipeline Stages for \DP Noise Injection}
\label{sec:stages}

As noted by~\cite{jayaraman2019evaluating}, there are three general positions in which \DP noise can be applied to a ML task, to preserve privacy of the data used, or the model built.
These three positions of entry in the ML pipeline are visualized in Figure~\ref{fig:eval_pipe}.
To make the next observations more concrete, let the function $\mathcal{F}$ map the training dataset \textbf{X} to class labels \textbf{y}, that is, $\mathcal{F}(\textbf{X}) = \textbf{y}$.
Then, the goal of the ML model is to approximately learn this relationship between dataset and labels as best as possible.
Next, we discuss each of these three Stages:

\noindent
\textbf{\textit{Stage 1 (S1):}} \textit{Before the learning process.}
During the collection or release of data (\textbf{X}), and before aggregation at the server, if local \DP noise is applied on every data record, the data (\textbf{X'}) are protected before being used in a ML pipeline.
Alternatively, when releasing a dataset to the public domain, the owner can train a data generator to create a synthetic dataset containing the same data semantics as the real data, but with the synthetic data governed by the rules of \DP.
Consequently, the $\mathcal{F}$(\textbf{X'}) model learned is \DP-enabled.

\noindent
\textbf{\textit{Stage 2 (S2):}} \textit{During the learning process.}
In this stage, each step of the model update is restricted as to not excessively alter the model with the added \DP noise ($\mathcal{F'}$(\textbf{X})), and thus compromise the privacy of a given batch of records.
The classic example for this Stage is the Tensorflow Privacy, which deploys a \DP stochastic gradient descent algorithm~\cite{abadi2016deep}.

\noindent
\textbf{\textit{Stage 3 (S3):}} \textit{After the learning process.}
After the data modeling has finished, the learned parameters of the model can be perturbed, by adding \DP noise on them ($\mathcal{F'}$(\textbf{X})) to remove dependencies between learned parameters and training data.

\subsection{\DP-based ML Algorithms}
\label{sec:algorithms}

A literature review on existing ML methods that provide \DP protection to the data or model revealed that various realizations can be loosely divided in the three Stages outlined above.
We identified two key ML classification algorithms of interest: Naive Bayes (\NB), and Neural Networks (\NN).
Next, we describe the approach used by each one in learning on data in a supervised setting, while applying \DP noise in each Stage.
We remark that~\cite{jayaraman2019evaluating} focused primarily on different \DP compositions for \NN and \LR, both leveraging empirical risk minimization in the learning process, and loosely mapping to our S2 and S3 Stages, respectively.
However, they did not offer direct comparisons of these ML algorithms across all Stages, as we do.
In fact, we compare these and other \DP-based ML methods, summarized in Table~\ref{tab:exec_dp_ways}, as applicable in each Stage.

\begin{table}[ht]
\vspace{-0mm}
\caption{\DP-enabled ML methods used in each pipeline Stage.}
\label{tab:exec_dp_ways}
\vspace{-3mm}
\centering
\resizebox{0.7\columnwidth}{!}{%
\begin{tabular}{p{3cm}p{1.0cm}p{1.0cm}p{0.8cm}}
\toprule
				 \multicolumn{4}{r}{Stage where \DP noise is applied}	\\
ML Method		&	S1	&	S2	&	S3			\\
\midrule
\rowcolor[HTML]{EFEFEF}
Naive Bayes		&	X	&		&	X			\\
\rowcolor[HTML]{EFEFEF}
Neural Network		&	X	&	X	&				\\
\bottomrule
\end{tabular}
}
\vspace{-2mm}
\end{table}

\subsubsection{S1: Manipulation of Laplacian Noise}

At this Stage, we apply \DP Laplacian noise~\cite{das2016tracking} directly on the dataset, and thus, this process is independent of the ML algorithm used in subsequent steps of the framework.
As a result of this independence, we can employ both ML algorithms, in their non-private versions, on the modified, \DP data.
In particular, \DP is provided by the addition of noise to every vector in the dataset.
Laplacian noise is independently sampled for every feature value, of every data vector from the distribution $Lap(0, \beta_i)$, where $\beta_i = \frac{S_i}{\epsilon/p}$, and $S_i$ is the value range of the $i$th feature~\cite{das2016tracking} (Algorithm~\ref{alg:dp-data}).

\begin{algorithm}[ht]
	\SetAlgoLined
	\KwIn{Training Dataset \textbf{$X$}, where $x := \{x_0, ..., x_i, ...,x_p\}$}
	\KwResult{Differentially private Training Dataset \textbf{$X'$}}

	\For{$x$ in $X$}{
		$x' = x + b;$ 
		where $b:= \{b_0, ..., b_i, ...,b_p\}; b_i \in Lap(0, \beta_i)$
	}

	Proceed with learning task $\mathcal{F}$ on \textbf{$X'$}.
\vspace{1mm}
\caption{Direct addition of \DP noise to dataset before ML.}
\label{alg:dp-data}
\end{algorithm}

Adding \DP noise in S1 is ML independent and permits more flexibility, as any ML or DL method can be employed after S1 for training.
The application of noise is dependent on data types and their complexity with respect to features and values allowed. 

\begin{remark}
\DP noise is applied with the assumption that features are independent from each other, meaning a maximal amount of noise must be applied to each feature to ensure \DP. With knowledge of feature dependence, hypothetically less noise can be applied to the dependent features as there is less uniquely identifying information between the dependent features.
\end{remark}

\subsubsection{S2: \DP-based Neural Networks}

Neural Networks (\NN) are designed to mimic the functionality observed within brains~\cite{gurney2014introduction}.
They contain multiple layers of neurons (some hidden) that are activated depending on the activation of neurons in the previous layer. 
The influence of a previous layer's neurons on the current neuron varies depending on a weight or parameter value learned during the training phase.
The very final layer is often a decision layer that corresponds to each of the classes present in the classification problem.
The degree of activation of this last layer is analogous to the \textit{confidence} of the class prediction.

The approach employed by Tensorflow-Privacy's~\cite{abadi2016deep} implementation of \DP-enabled \NN involves the use of a \DP stochastic gradient decent (SGD) algorithm.
The SGD algorithm seeks to find network parameters $\theta$ to learn function $\mathcal{F}$.
The \DP-based SGD first clips or limits the size of gradient update, to not be heavily impacted by one batch of data.
Additional noise is added to the updated gradient depending on the values of $\epsilon$, and batch sensitivity (Algorithm~\ref{alg:dp-nn}):

\begin{algorithm}[ht]
	\SetAlgoLined
	\KwIn{Training Dataset \textbf{$X$}}
	\KwResult{Differentially private parameters $\theta'$ }

	$\theta' \leftarrow \texttt{RAND}$, initialize the parameters randomly;

	\For{batch $t \in T$}{
		compute gradient $\Delta \theta$,
		clip gradients $\Delta \theta$,
		add DP-noise $b$
		
		$\theta' = \theta' + \Delta\theta + b$
	}
	Complete $\mathcal{F}'$ learning task with $\theta'$.
\vspace{1mm}
\caption{\DP-based Stochastic Gradient Decent~\cite{abadi2016deep}.}
\label{alg:dp-nn}
\end{algorithm}

\subsubsection{S3: \DP-based Naive Bayes}

The Naive Bayes (\NB)~\cite{rish2001empirical} classification algorithm learns probabilistic distributions of the output classes informed by the input feature values.
The algorithm is considered ``naive'', as it assumes an independence between features.
The distributions are learned directly from the training dataset.
The simple formulation of the model enables the Naive Bayes classifiers to both be trained, and to make predictions relatively quickly.

IBM \NB~\cite{holohan2019diffprivlib} implements an $(\epsilon)$-\DP \NB, originally by~\cite{vaidya2013differentially}.
The approach adds noise to the learned distributions that relate the input feature to the output decision.
Algorithm~\ref{alg:dp-nb} shows the Laplacian noise addition to the mean and standard deviation ($\mu$, $\sigma$) computed from training dataset $X$.
A more complete algorithm for handling both categorical and continuous data can be found in~\cite{vaidya2013differentially}.

\begin{algorithm}[ht]
	\SetAlgoLined
	\KwIn{Training Dataset \textbf{$X$}}
	\KwResult{Differentially private model distributions \textbf{$\theta'$}}
	Compute (\textbf{$\mu$, \textbf{$\sigma$}}) from \textbf{$X$};
    
	\For{$i$ in $features$}{
		Compute scaling factor $S_{(\mu,i)}$ and $S_{(\sigma,i)}$ from feature mean $\mu_i$, feature STD $\sigma_i$, and $\epsilon$;

		$\mu_i' = \mu_i + b_i$;
		where $b_i \in Lap(0, S_{(\mu,i)})$;
    
		$\sigma_i' = \sigma_i + b_i$;
		where $b_i \in Lap(0, S_{(\sigma,i)})$;
	}
	Compute output priors $P(y|x)$ from (\textbf{$\mu'$, \textbf{$\sigma'$}});
\vspace{1mm}
\caption{\DP-based Naive Bayes provided by IBM~\cite{holohan2019diffprivlib}.}
\label{alg:dp-nb}
\end{algorithm}

\subsection{Privacy Attacks \& Privacy Metrics}
\label{sec:privacy-attacks}

Traditionally, privacy has been measured with theoretical metrics such as information leakage~\cite{m2012measuring, issa2016operational} and mutual information~\cite{wang2016relation}.
However, recent privacy attacks such as membership inference (MI)~\cite{shokri2017membership, salem2018ml, yeom2018privacy} and attribute inference (AI)~\cite{yeom2018privacy, zhao2019inferring} have been introduced~\cite{jayaraman2019evaluating} as alternatives to measure privacy risk of ML models.

In this work, we quantify the privacy offered by the implementation of \DP, through the effectiveness of these two well-known privacy attacks (MI and AI).
The threat model adopted by these attacks falls under the category of black-box attacks, with an adversary only having access to the input and output of the ML model (though white-box approaches can enhance the attack performance~\cite{nasr2019comprehensive}).
In fact, for the current generation of MI attacks~\cite{shokri2017membership, yeom2018privacy,salem2018ml}, only one query is required for the vector in question (disregarding queries needed to train an attack model), whereas AI attacks need multiple queries, one for any possible value in the unknown attribute.

Next, we survey multiple MI and AI attacks.
However, in the experimental part of this work we focus on the MI attack of \SalemMI and the AI attack of \YeomAI (see details next).

\subsubsection{Membership Inference Attack}

\textit{MI attack}~\cite{shokri2017membership,salem2018ml,yeom2018privacy} defines an attacker that tries to determine if a specific data record has been included within the training data of a given ML model, or not.
The attack objective is related to the definition of \DP, as according to \DP, two datasets with or without an $\epsilon$ proportion of records should be indistinguishable from each other.
Of course, this is problematic if a privacy ML practitioner is seeking to maintain the confidentiality of their training data, or to adhere to privacy regulations governing the data used in training.
In literature, there are three realizations of the MI attack~\cite{shokri2017membership, salem2018ml, yeom2018privacy}.

\noindent
\textbf{\SalemMI~\cite{salem2018ml}} attack works on the premise that a ML model is more confident about a prediction on an input vector it has previously encountered (in the training set), than an input vector it has not previously encountered (in the testing set).
Thus, a vector with a higher prediction confidence on any class label is more likely to be a member vector.
A threshold can be found from a similarly distributed dataset to make a final distinction if an input is a member or non-member.
Indeed, this attacker does not know the vector's classification truth, and the prediction confidence is a single value of the most probable class, irrespective of if it is the correct prediction.

\noindent
\textbf{\YeomMI~\cite{yeom2018privacy}} attack is similar to \SalemMI.
However, they use prediction loss, requiring the true label of the input vector.
Additionally, instead of finding a threshold from a similar data distribution, the model training loss is assumed known and used as the threshold.
The additional information needed makes the \YeomMI attack more difficult to mount than \SalemMI, but more effective.

\noindent
\textbf{\ShokriMI~\cite{shokri2017membership}} Shokri's MI attack trains shadow models that replicate the behavior of the target model, from which an attack model is trained to differentiate between members and non-member vectors from the training and testing process of said shadow models. The attack model takes as an input, the prediction probabilities of all classes for a given vector. The shadow models allow an attacker to produce larger datasets for the attack model, and thus train a superior attack model.
However, the process of training shadow models is a computation and data intensive operation (Though~\cite{salem2018ml} demonstrates only one shadow model is required, in addition to an ability to train an attack model on a different dataset and transfer said attack model to the target model and data).

\subsubsection{Attribute Inference Attack}
\textit{AI attack} is an extension of the MI attack, however, instead of only determining if a record is included within the training set, the adversary seeks to recover the exact value of a missing attribute that could be masked due to it's sensitivity (e.g., the diagnosis for type of cancer of an individual).
In particular, if a record vector has a dimensionality of $n$ (i.e., $n$ features), the adversary is assumed to have $n-1$ true features of the original record.
Their objective is to infer the $n^{th}$ feature's sensitive value.
In general, AI attacks are more difficult to mount than MI attacks due to the requirements on the attacker.

The first method of AI \textbf{(\YeomAI)} follows work by~\cite{yeom2018privacy} and~\cite{jayaraman2019evaluating} in evaluating every binned permutation of a vector and its unknown attribute, and selecting the value that produces a loss closest to the model's training loss.
The second attack \textbf{(\SalemAI)} follows work by \cite{zhao2019inferring} and \cite{salem2018ml}, by selecting the vector permutation that produces the highest model confidence as the most likely real attribute. 

To date, many implementations of AI attack (e.g., as in~\cite{jayaraman2019evaluating}) bin numerical features for a binary evaluation.
In this work, we go beyond the state-of-art and increase the number of allowable (binned) values in the inference of a vector's attribute, from two bins up to a maximum to 10 bins, depending on the unique values of an attribute.
For instance, if an attribute is binary, two bins are required.
A numerical feature with 6 distinct values will require 6 bins, and a continuous feature will be binned into 10 value bins.

\subsubsection{Measuring Privacy Leaks: Adversary Advantage}

The \textit{adversary advantage} can be described as the improvement of a privacy attack observed on a set of input vectors that were included in the training set, as opposed to not being included in the training set.
The rate at which the privacy attack succeeds on the positive class (member vectors) is the True Positive Rate (TPR), while the rate at which privacy attack is incorrectly predicted on the negative class (non-members) is the False Positive Rate (FPR).
As such, the advantage can be formulated as \textbf{$ADV = TPR - FPR$}.
A rigorous definition of the advantage is provided in~\cite{yeom2018privacy}.
It is clear novel attacks, and their advantage can be added in our framework. Here, we shall measure the impact of the \SalemMI and \YeomAI attacks.

\subsection{ML Utility Metrics}
\label{sec:utility-metrics}

The objective of ML is to learn trends from a training dataset, and then predict the label of a previously unseen input instance.
To evaluate the effectiveness of a trained model, predictions are made on a holdout set (not used in training), said predictions are then compared to known true labels.
The proportion of the holdout set that is correctly re-predicted as the true labels, represents the accuracy (ACC) of the trained model: $ACC = n_{correct}/n_{holdout}$.

Accuracy is a simple measure of ML prediction performance.
Other commonly used metrics are AUC, Precision, Recall, or F-Score.
Also, new metrics such as model fairness~\cite{corbett2018measure} and minimization of computational processes~\cite{bottou2010large, chauhan2018breathing} can be important in a privacy-utility tradeoff.
All such utility metrics can be added in our framework.

We focus on \textbf{Accuracy Loss (\ACL)}, defined as the ratio of performance lost when \DP is applied to the ML process ($m$), in comparison to an equivalent ML model trained with no \DP applied (i.e., $\epsilon = \inf$):
\vspace{-3mm}
\begin{equation}
\texttt{Accuracy Loss (\ACL)} = 1 - \frac{ACC_{(m, \epsilon)}}{ACC_{(m, \epsilon=\inf)}}
\end{equation}

\section{EXPERIMENTAL INVESTIGATION}\label{sec:experiments}

In this section, we detail how the methodology introduced earlier is instantiated\footnote{We provide our code and data at \url{https://github.com/PrivateUtility/PrivateUtility}} to experimentally investigate the tradeoff between ML model performance with respect to prediction, vs. privacy guarantees provided to data used to train said model.
In particular, with our experimentation, we are interested in answering questions of:
\begin{enumerate}
\itemsep0em
\item What is the inflection point in the tradeoff between ML model accuracy and privacy leak? Is this inflection point consistent across various types of privacy attacks?
\item Does the stage of the \DP-enabled ML framework in which the \DP noise is applied impact this inflection point?
\item Is there a ML method that outperforms others at both prediction and privacy guarantees, consistently across datasets?
\end{enumerate}

We seek to empirically identify important parameters that affect the manifestation of this privacy-utility tradeoff.
To this end, in Sec.~\ref{sec:exp-setup}, we detail the experimental procedures that vary the \DP noise amount ($\epsilon$), where it is applied in the framework (Stages), different \DP-ML algorithms implemented and metrics used.
Then, we describe the training datasets used, both synthetic and real (Sec.~\ref{sec:datasets}), providing details on number of classes and type of attributes (continuous, binary).
In the next Section~\ref{sec:exp-results}, we present our experimental results and extract key takeaway messages.

\subsection{Experimental Framework}\label{sec:exp-setup}

First, we detail implementations of \DP-ML methods used, as well as metrics to assess ML performance and privacy when \DP noise is applied.
We note that Sec.~\ref{sec:methodology} already provided details for the privacy attacks and ML methods used.
Then, we outline the common steps shared between all evaluations of the \DP-ML methods. 
We bootstrapped our framework implementation from~\cite{usenixsecGithub2019-dp}, but make the following crucial extensions:
\begin{squishlist}
\item accommodate the new ML algorithms to run in this framework,
\item adapt code to improve framework resource consumption,
\item add implementation of MI attack proposed by Salem et al.~\cite{salem2018ml},
\item adapt AI attack of Yeom et al.~\cite{yeom2018privacy} to support multiple bin values instead of only binary,
\item add synthetic data generation for tradeoff \& benchmark studies.
\end{squishlist}

\subsubsection{Machine \& Deep Learning Methods}

We used implementations of ML algorithms explained in Sec.~\ref{sec:algorithms} readily available online.
\textit{Tensorflow-Privacy}~\cite{abadi2016deep} has code in~\cite{nn-dp}.
\textit{IBM Naive Bayes}~\cite{holohan2019diffprivlib} has code in~\cite{nb-lr-dp}.
The hyper-parameters of the \NN models were replicated from~\cite{jayaraman2019evaluating}.
All other models' parameters are kept at library defaults.

\subsubsection{Performance Metrics \& Privacy Budget}

In our experiments with the various ML methods and datasets, we measure different performance metrics.
For prediction performance of a trained model, we measure Accuracy Loss (\ACL) (See Sec.~\ref{sec:utility-metrics}), 
We perform two MI and AI attacks, \SalemMI and \YeomAI (See Sec.~\ref{sec:privacy-attacks}), to quantify privacy leaks.
Finally, in order to vary the amount of \DP noise applied in each framework Stage and in each ML method, we use different values for the privacy budget: \\
$\epsilon = \{ 0.01, 0.05, 0.1, 0.5, 1, 5, 10, 50, 100, 500, 1000 \}$.

\subsubsection{Experimental Steps}

To perform the evaluation for: 1) a given dataset, on 2) a \DP-based ML method, with 3) a privacy budget $\epsilon$, we first sample from the dataset two sets of 10,000 samples each, forming our training and testing sets.
Then, we train the ML model with the training set.
In the case of S1 \DP noise, we apply noise to the training set prior to the model training. 

Each model's prediction accuracy is obtained on the unseen testing set.
With a trained model, the \SalemMI, and \YeomAI attacks are performed.

In MI attacks, the training set constitutes the membership set, whilst the testing set is the non-member test set.
In AI attacks, we consider up to 10 unique values for the unknown protected attribute (whilst accounting for continuous features).
The attack is repeated on 20 different attributes, randomly selected to be the protected attribute.
Then, the entire training and attack process is repeated 5 (10) times for synthetic (real) data, with training and testing sets sampled anew, to reduce impact of biases arising from the data or \DP noise.

\subsection{Experimental Datasets}\label{sec:datasets}

\subsubsection{Synthetic Data}\label{sec:synthetic-data}

We generated data by uniformly sampling $100k$ vectors from a normalized feature space of $50$ features.
From these $100k$ vectors, we apply k-means clustering onto the dataset to artificially create labels of 2, 5, 10, 20, 50, 100 and 200 classes. This results in 7 different datasets of varying number of classes, however, they all contain the same vectors originally sampled.
 
\subsubsection{Real-World Data}\label{sec:real-data}

We used three  real datasets to study the tradeoff in our \DP-enabled framework (summary in Table~\ref{tab:dataset-summaries}):

\noindent
\textbf{CIFAR-100~\cite{krizhevsky2009learning}:}
The \textit{CIFAR} dataset consists of $50k$ tiny images of various objects, that can be labeled according to $100$ types.
They can also be re-classified under $20$ type super-classes.
This dataset has been pre-processed with principal component analysis as in~\cite{jayaraman2019evaluating}, to extract $50$ key features to represent each of the images.

\noindent
\textbf{Purchase~\cite{purchase_dataset}:}
The \textit{Purchase} dataset contains $200k$ user records of item purchases made from a set of 599 products.
The values are binary, indicating if users had or not bought one of the 599 items.
We perform a similar pre-processing step as in~\cite{shokri2017membership}, by encoding of a single user's transaction history as a binary vector, followed by the k-means clustering of users into purchaser groups.
We consider label complexities of $k=\{2, 10, 20, 50, 100\}$.

\noindent
\textbf{Netflix Prize~\cite{netflix_dataset}:}
The \textit{Netflix} dataset was first released in 2006, and contains ratings (from 1 to 5) by viewers on the Netflix platform for movies they watched.
This dataset was also used in~\cite{yeom2018privacy}.
However, insufficient pre-processing details were provided for us to replicate their exact dataset.
Therefore, we performed the following steps:
(1) Sample the user ratings of the top 1000 rated (based on number of ratings, not rating score) movies within the dataset.
(2) Every user has its ratings assembled into a feature vector, with unrated movies filled in with a zero value.
(3) If a user has not rated any of the 1000 most popular movies, the user is excluded from the dataset.
(4) Then, we apply k-means clustering (as in \textit{Purchase}) to obtain viewer groupings of $k= \{2, 10, 20, 50, 100\}$.

\begin{table}[t]
\caption{Summary of datasets used in our experimental investigation, with respect to number of instances available, classes provided (or constructed), and attributes available.}
\label{tab:dataset-summaries}
\vspace{-3mm}
\centering
\resizebox{0.90\columnwidth}{!}{%
\begin{tabular}{llll}
\toprule
\textbf{Dataset}		&	\textbf{Instances}	&	\textbf{Classes}		&	\textbf{Attributes}	\\
\midrule
Synthetic		&	100,000		&	2, 5, 10, 20, 50, 100, 200		&	50		\\
\midrule
CIFAR~\cite{krizhevsky2009learning}		&	50,000		&	20, 100		&	50		\\
\midrule
Purchase~\cite{purchase_dataset}		&	200,000		&	2, 10, 20, 50, 100		&	599	\\
\midrule
Netflix~\cite{netflix_dataset}			&	100,000		&	2, 10, 20, 50, 100		&	1000	\\
\bottomrule
\end{tabular}
}
\end{table}


\section{Experimental Results}
\label{sec:exp-results}

In this section, we present our results for different experiments using our evaluation framework, to answer the questions posed in Sec.~\ref{sec:experiments}.
We first analyze results with synthetic data, while controlling class complexity, and extract generalized patterns related to the privacy-utility tradeoff.
We shall compare these patterns with results on real data to assess how the tradeoff manifests on the real-world datasets.

\subsection{Privacy-Utility Tradeoff on Synthetic Data}
\label{sec:synth-result}

We perform experiments on controlled, synthetic datasets to discover generalizable properties that can be drawn regarding the privacy-utility tradeoff.
The synthetic dataset allows us to remove the effect of data-specific biases (in a controlled manner), that may otherwise be present in the real data.

\subsubsection{ML accuracy vs. \DP noise}

\begin{figure*}[t]
\begin{center}
\subfigure[S1: Naive Bayes]{\includegraphics[width=0.48\columnwidth]{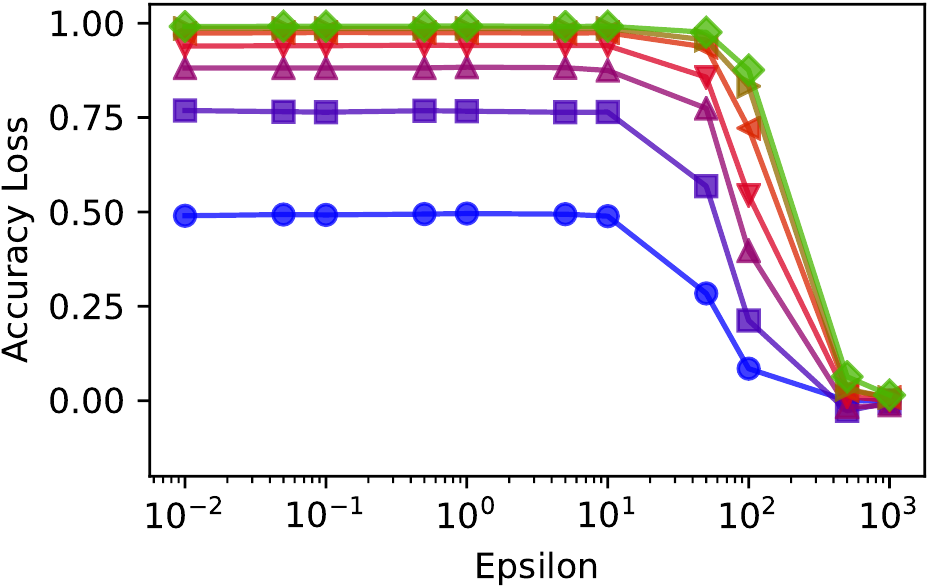}}
\hspace{2pt}
\subfigure[S1: Neural Network]{\includegraphics[width=0.48\columnwidth]{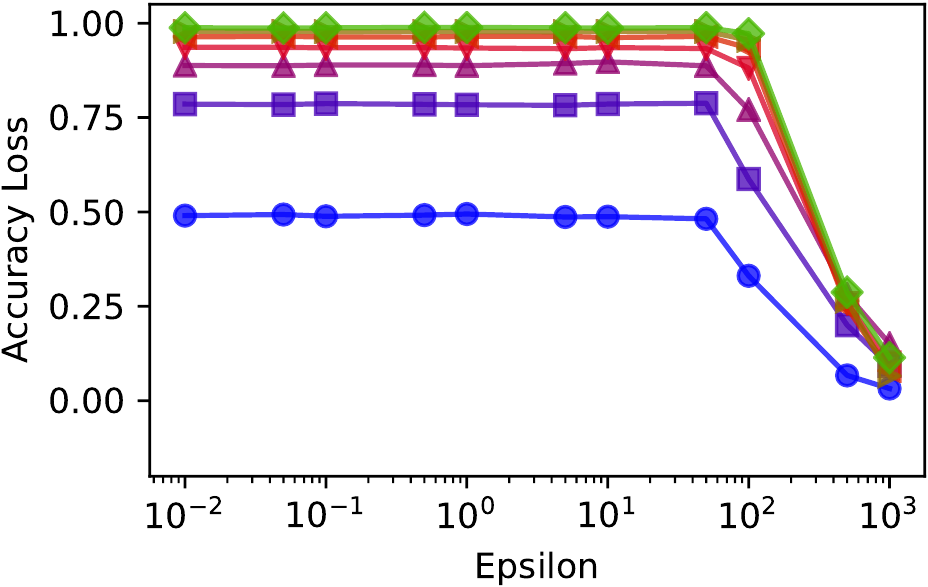}}%
\hspace{2pt}
\subfigure[S2: Neural Network]{\includegraphics[width=0.48\columnwidth]{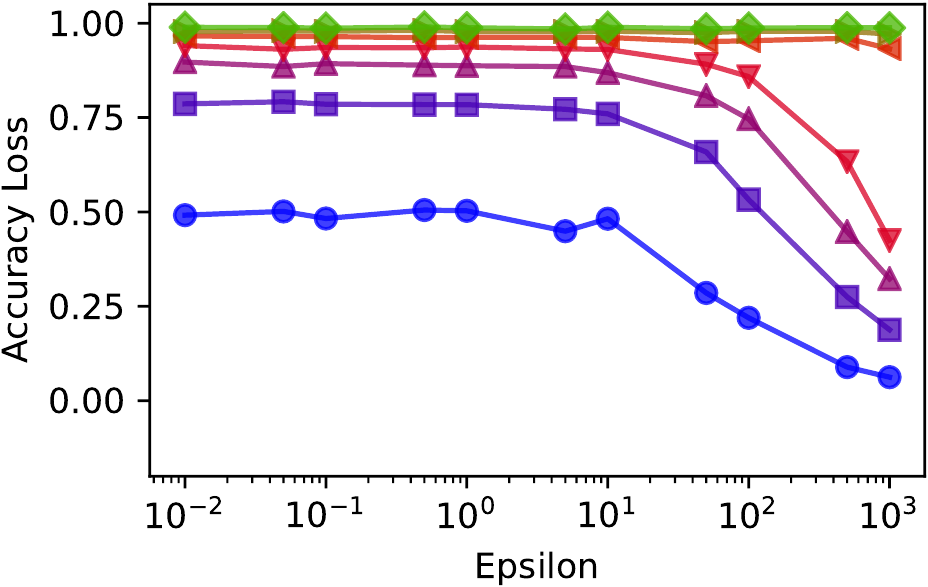}}
\hspace{2pt}
\subfigure[S3: Naive Bayes]{\includegraphics[width=0.48\columnwidth]{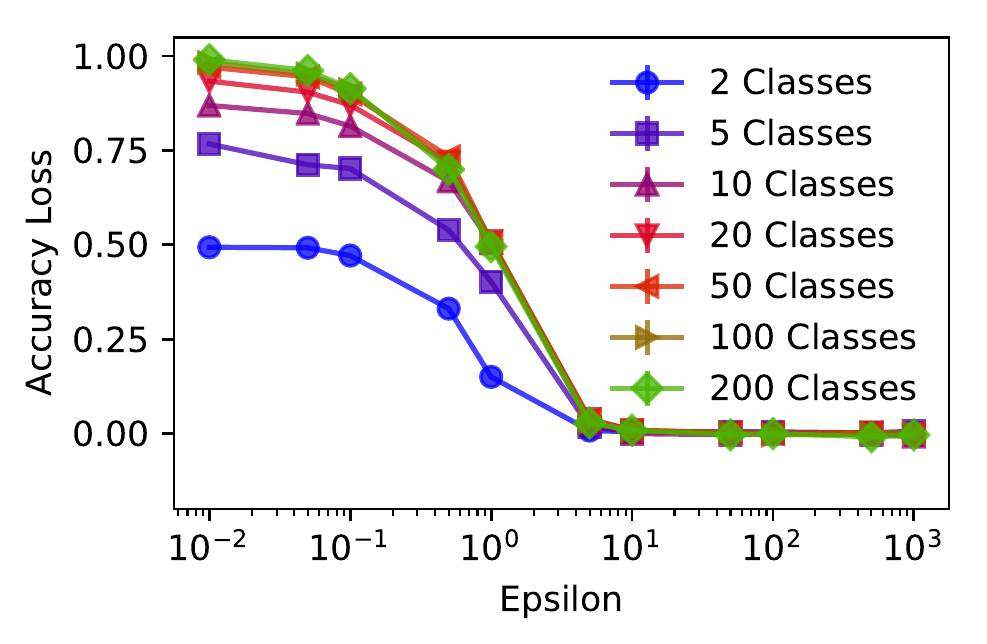}}
\vspace{-5mm}
\caption{Accuracy Loss for each ML method used, when different amount of \DP noise is applied at framework Stages 1, 2 or 3, and for synthetic dataset complexities used. The underlying complexity of data vectors in each dataset remains the same.}
\label{fig:stageall-ACL-synth}
\vspace{-5mm}
\end{center}
\end{figure*}

In Figure~\ref{fig:stageall-ACL-synth}, we analyze \ACL and its inflection point for the different ML algorithms, while varying class complexity, amount of \DP noise and the stage at which it is applied.
When applying large amounts of \DP noise (i.e., small $\epsilon$) at the input Stage (Stage 1), we observe that the \ACL is equivalent to a random guess irrespective of the ML algorithm in use.
It is not until $\epsilon$=$10$ for \NB and $\sim$$100$ for \NN that the \ML algorithm is capable of outperforming a random guess.
In Stage 2, \NN exhibits a notable inflection point at $\epsilon$=$10$.
Finally, at Stage 3, the inflection point for \NB occurs at $\epsilon$=$0.01$.
When we compare ML performance across Stages, we observe that from S1 to S3, there is an increasing amount of \DP noise that can be applied to the ML method, before the accuracy of the system is reduced to a random guess.

Notably, given that the synthetic dataset is generated with the same underlying data vectors but with different class complexities, we observe that the inflection point occurs at about the same value of $\epsilon$, irrespective of the number of classes.

This inflection point does not vary across class complexities, but the complexity of each dataset does have a direct impact on the maximum \ACL (due to the random guess).

\begin{figure*}[t]
\begin{center}
\subfigure[S1: Naive Bayes]{\includegraphics[width=0.48\columnwidth]{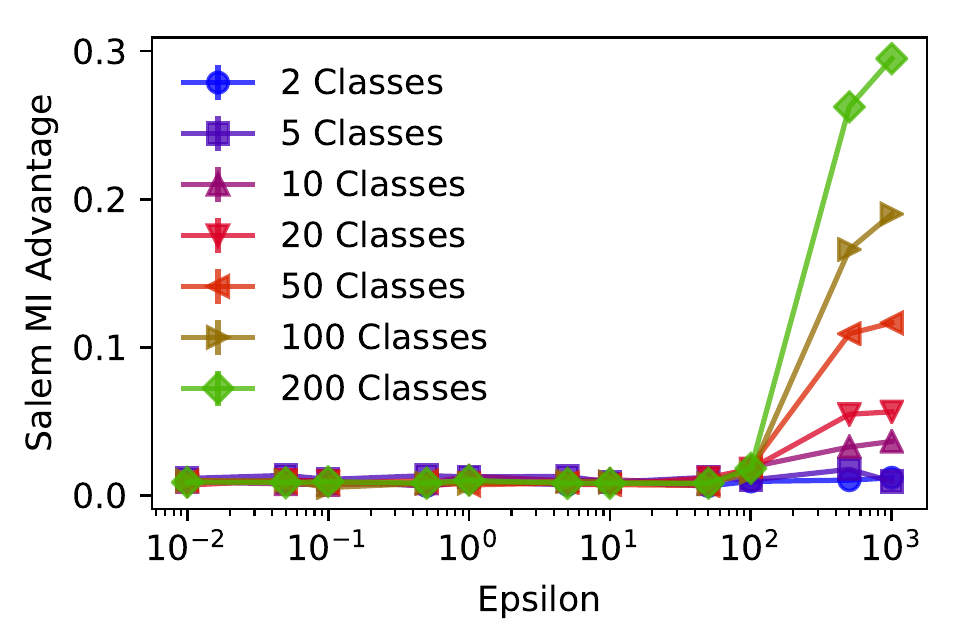}}
\hspace{2pt}
\subfigure[S1: Neural Network]{\includegraphics[width=0.48\columnwidth]{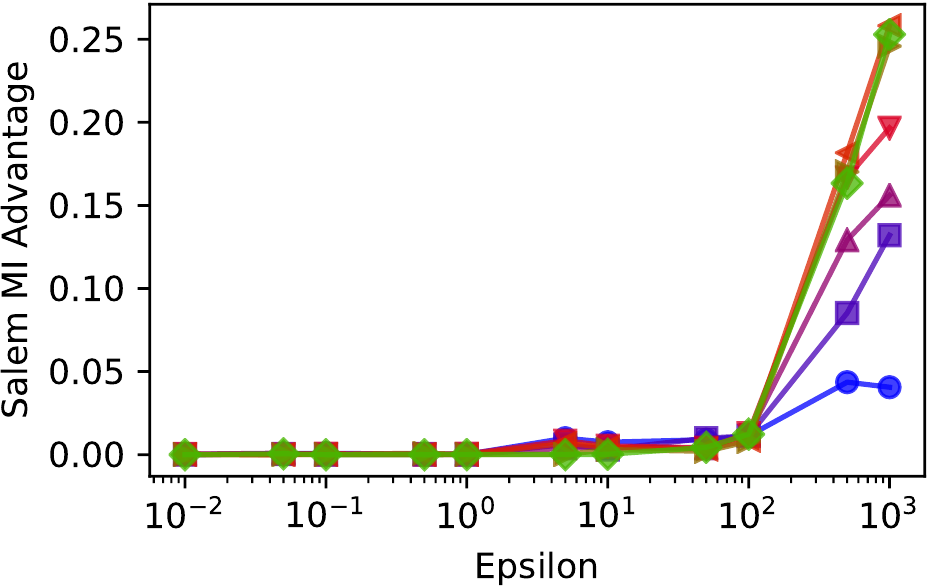}}%
\hspace{2pt}
\subfigure[S2: Neural Network]{\includegraphics[width=0.48\columnwidth]{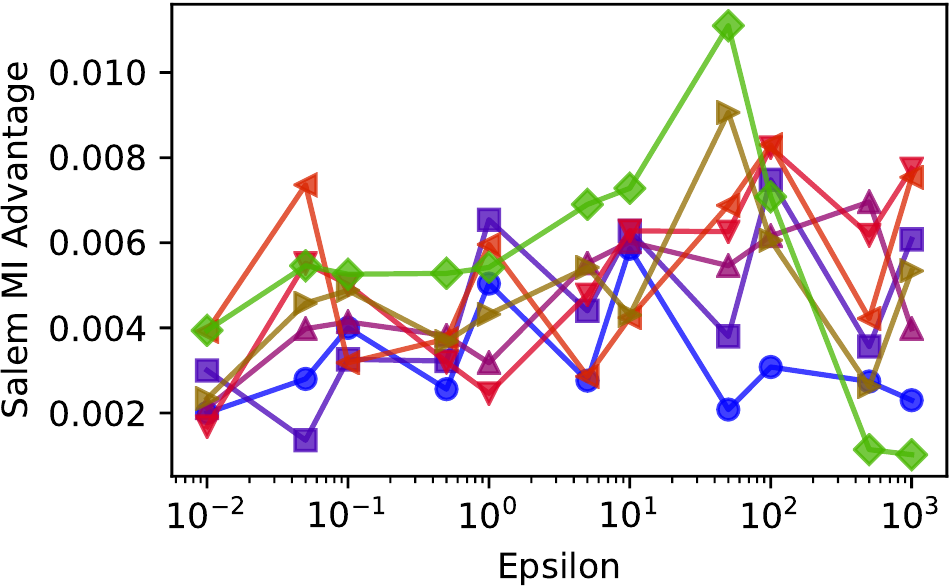}}
\hspace{2pt}
\subfigure[S3: Naive Bayes]{\includegraphics[width=0.48\columnwidth]{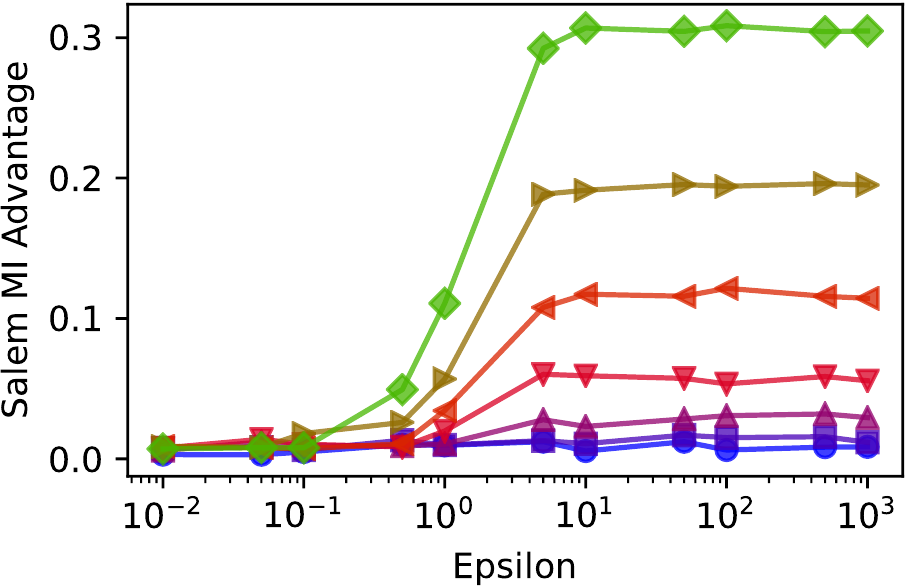}}
\vspace{-5mm}
\caption{Advantage of \SalemMI attack for each ML method, when different amount of \DP noise is applied at Stages 1, 2 or 3, and for different synthetic dataset complexities. The underlying complexity of data vectors in each dataset remains the same.}
\label{fig:stageall-SalemMI-synth}
\vspace{-5mm}
\end{center}
\end{figure*}

\begin{figure*}[t]
\begin{center}
\subfigure[S1: Naive Bayes]{\includegraphics[width=0.48\columnwidth]{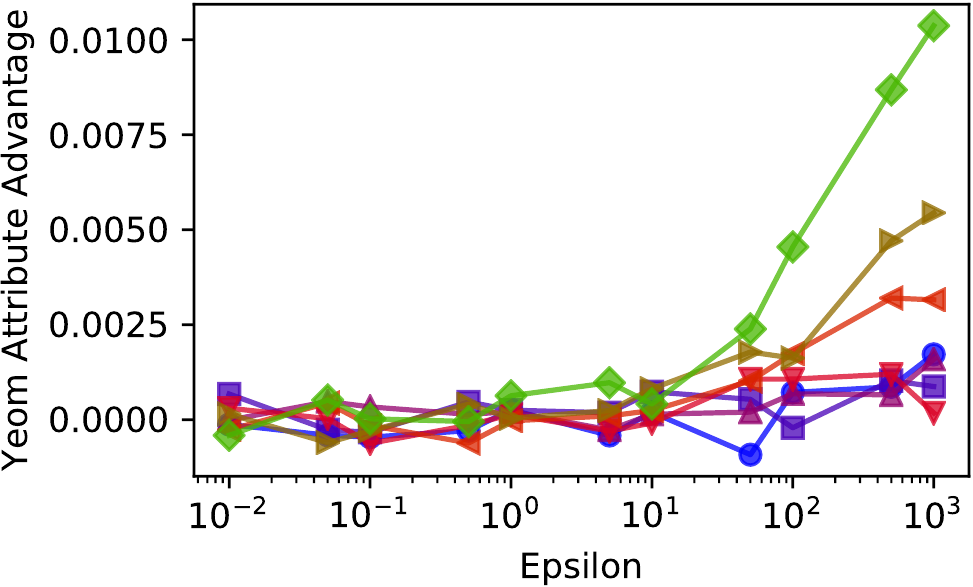}}
\hspace{2pt}
\subfigure[S1: Neural Network]{\includegraphics[width=0.48\columnwidth]{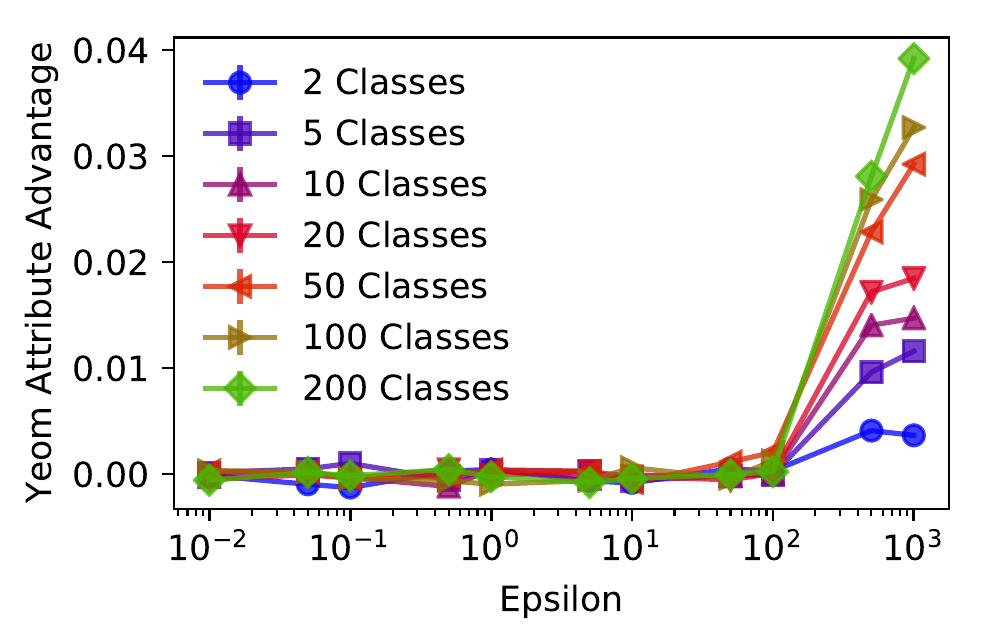}}%
\hspace{2pt}
\subfigure[S2: Neural Network]{\includegraphics[width=0.48\columnwidth]{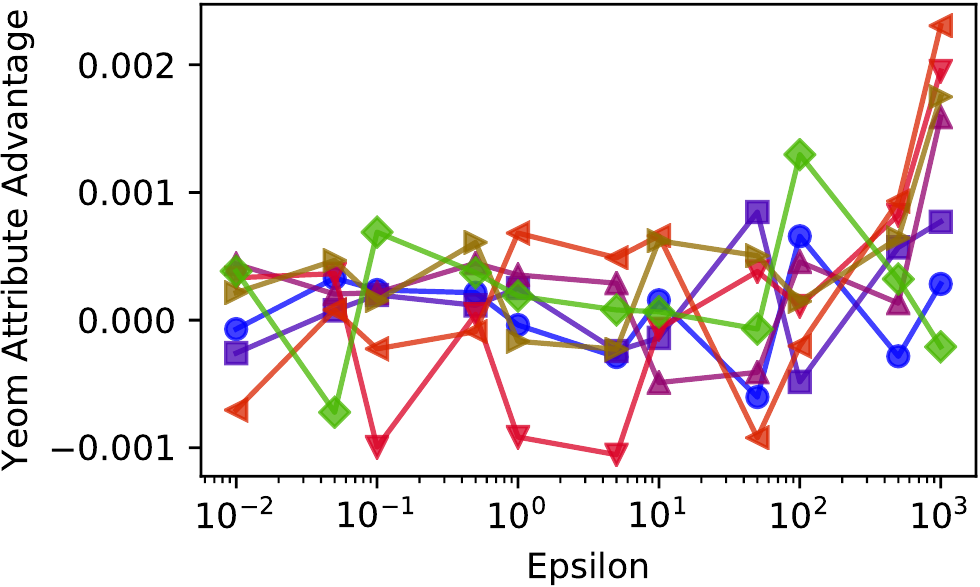}}
\hspace{2pt}
\subfigure[S3: Naive Bayes]{\includegraphics[width=0.48\columnwidth]{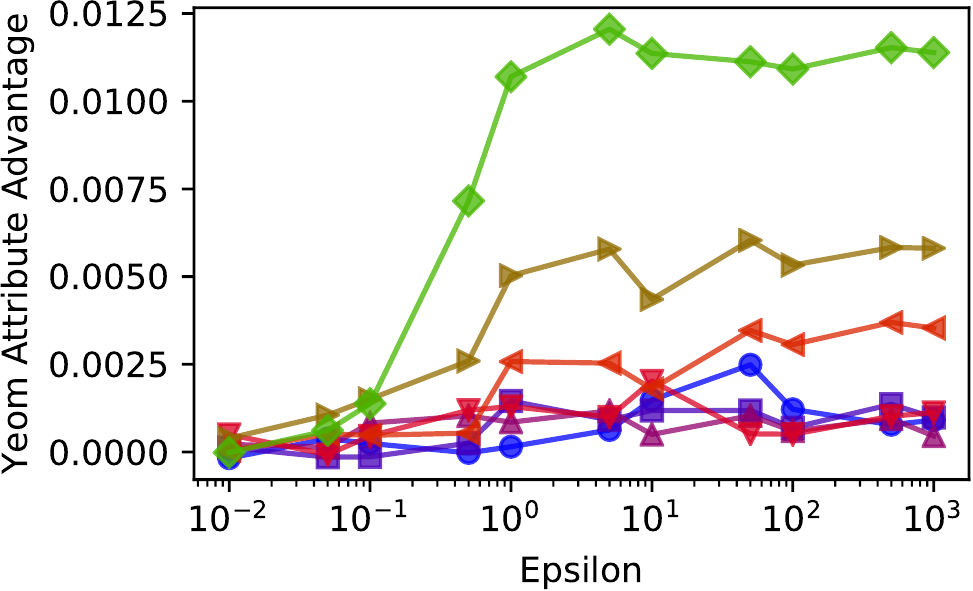}}

\vspace{-5mm}
\caption{Advantage of \YeomAI attack for each ML method, for different amount of \DP noise applied at Stage 1, 2, or 3, for different synthetic dataset complexities. The underlying complexity of data vectors in each dataset remains the same.}
\label{fig:stageall-YeomAI-synth}
\vspace{-4mm}
\end{center}
\end{figure*}

\subsubsection{Membership Inference Attacks vs. \DP noise}

In Figure~\ref{fig:stageall-SalemMI-synth}, we analyze the results on \SalemMI attack.
We first analyze the inflection point of the privacy advantage of the attacker, for each of the framework stages, followed by an analysis on the class complexity.

Across all ML methods in Stage 1, there is a clear inflection point at $\epsilon$=$100$, where an attacker until this point has a privacy advantage on vectors within the training data.
It is interesting to note that in comparison to the \ACL, the \SalemMI advantage reaches zero before the accuracy is completely diminished.
In Stage 2, the absolute advantage is rather small, resulting in a seemingly high variance.
Finally, in Stage 3, the inflection point for \NB occurs at $\epsilon$=$1$$\sim$$10$.
We note that the gradient of decreasing \SalemMI advantage (i.e., while $\epsilon$ is decreasing), is similar between S1 and S3, while the inflection points in S3 occur at smaller $\epsilon$ values than S1.

Across Stages, we note that for ML methods in S2, the \SalemMI struggles with very low advantages, in comparison to S1 and S3.
Similar to what was observed in \ACL, the class complexity appears to have little effect on the inflection point of the \SalemMI attack.
However, where the attack is effective, a higher class complexity is more vulnerable to \SalemMI attack.

\subsubsection{Attribute Inference Attacks vs. \DP noise}

In Figure~\ref{fig:stageall-YeomAI-synth}, we analyze the results on \YeomAI attack.
We study the inflection point of the attack for each stage and across stages, and across class complexity.

Across S1, there is an inflection point in the attack effectiveness at $\epsilon$=$10$$\sim$$100$.
We note that the absolute advantage of the attack differs depending on the ML algorithm used.
For S2, the \NN inflection occurs at $\epsilon$$\approx$$100$.
Lastly, for S3, \NB has inflection point at $\epsilon$=$0.1$.

The peak of the Yeom AI attack does not appear to perform well on S2:\NN relative to the other ML methods, however it is noted that the ACL for this model did not approach zero like the other 3 models as previously seen in Figure~\ref{fig:stageall-ACL-synth}, though for \NB between S1 and S3, the absolute AI advantage remains relatively similar.
However, adding noise at stage S3 seems to have a bigger impact on the inflection points of \NB than when adding noise at Stage S1.

Again, we observe that number of classes does not impact the position of the inflection point for a given DP ML technique.

\subsubsection{Summary of Results on Synthetic Data}

We saw evidence of a measurable inflection point and the tradeoff between the utility and privacy, as measured by MI and AI attacks.
The observable $\epsilon$ value in which this inflection occurs, is largely dependent on the Stage in which the \DP noise is applied, and to a lesser extent on the algorithm used.
Between \ACL and privacy advantage results, class complexity has little impact on where the tradeoff is observed.


\subsection{Privacy-Utility Tradeoff on Real Data}
\label{sec:real_data}

Here, we present an analysis of results on real data, highlighting pattern similarities and differences compared to the synthetic data.

\subsubsection{ML accuracy vs. \DP noise}
\label{sec:acl-results}

Next, we analyze the \ACL on real data in a similar fashion as with the synthetic data, 
but grouping results of all datasets by class complexity (number of classes) to facilitate comparison.
We discuss model performance at each stage and across stages, and the impact of class complexity on \ACL.

\begin{figure*}[t]
\begin{center}
\subfigure[S1: Naive Bayes]{\includegraphics[width=0.48\columnwidth]{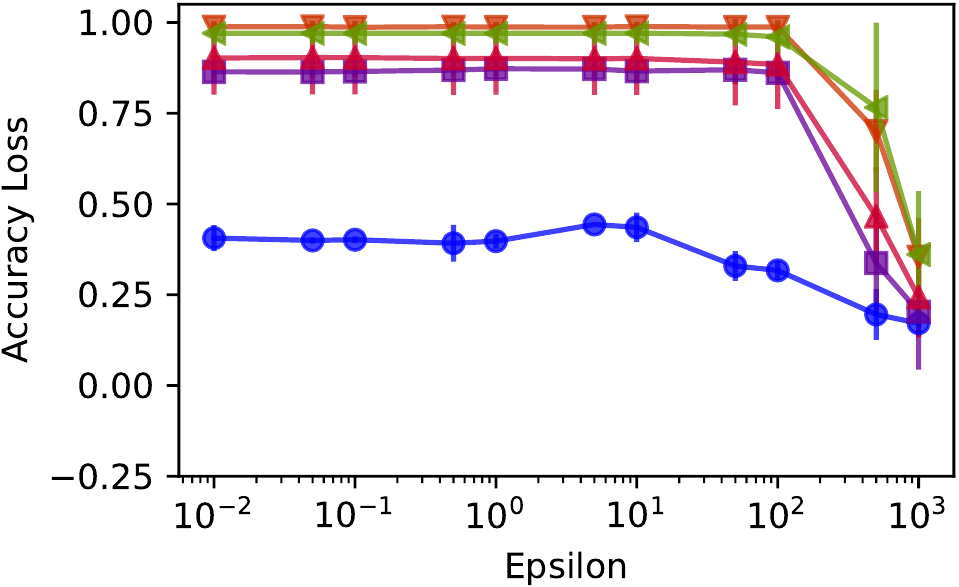}}
\hspace{2pt}
\subfigure[S1: Neural Network]{\includegraphics[width=0.48\columnwidth]{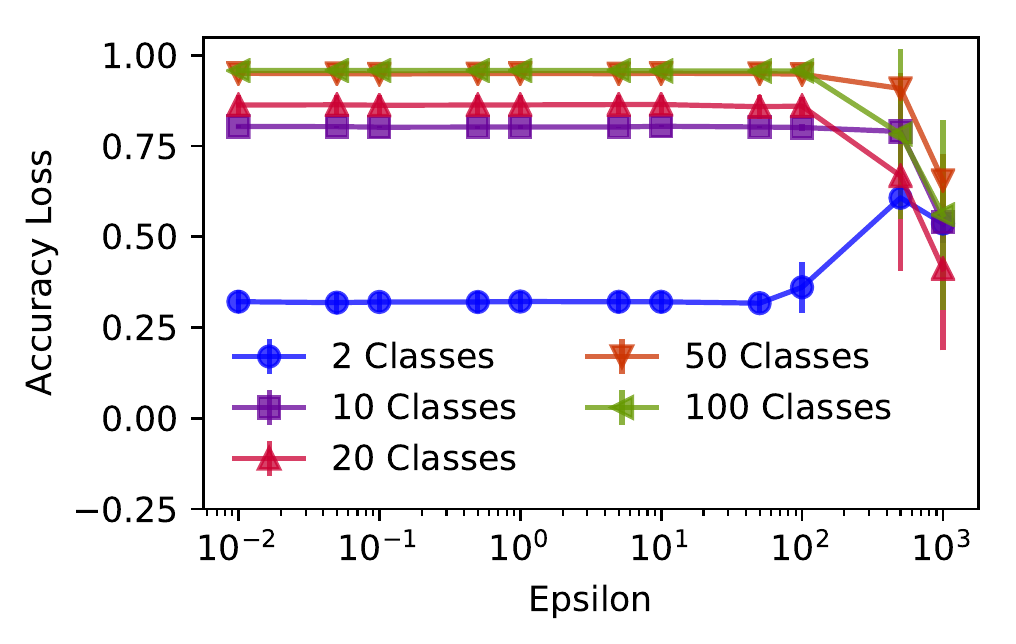}}%
\hspace{2pt}
\subfigure[S2: Neural Network]{\includegraphics[width=0.48\columnwidth]{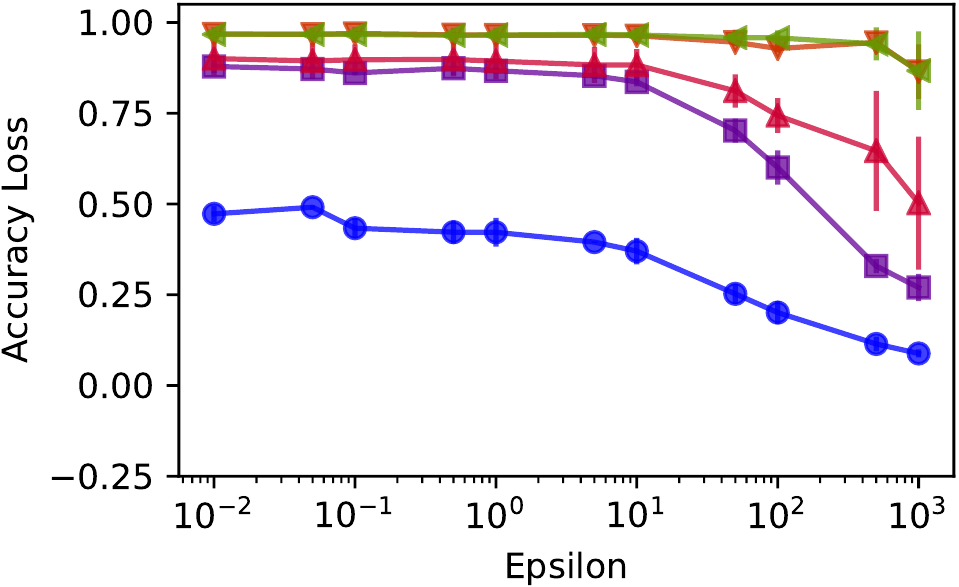}}
\hspace{2pt}
\subfigure[S3: Naive Bayes]{\includegraphics[width=0.48\columnwidth]{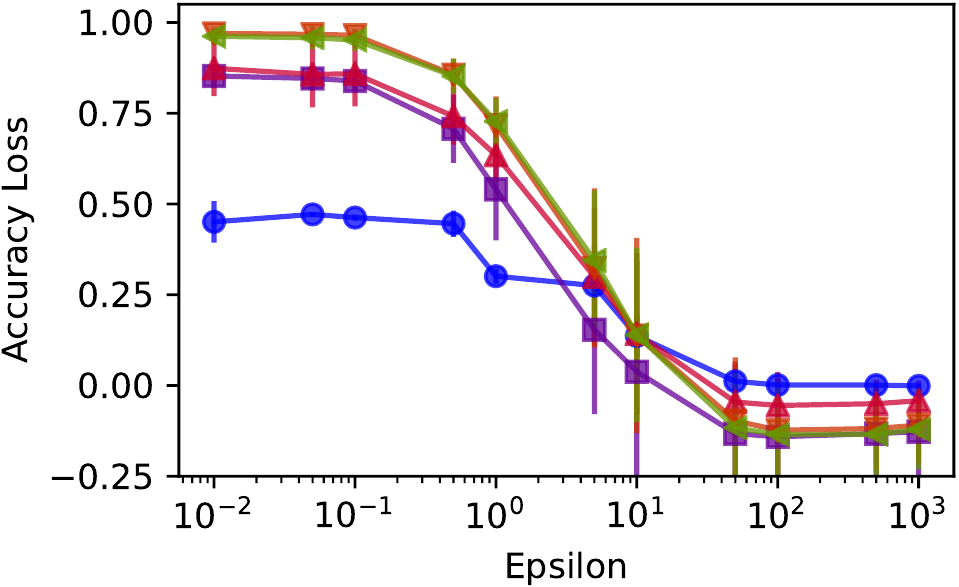}}
\vspace{-5mm}
\caption{Accuracy Loss for each of the ML methods used, when different amount of \DP noise is applied at Stage 1, 2 or 3 of the framework, and for different real datasets used. We summarize the datasets by the number of classes used.}
\label{fig:stageall-ACL-all}
\vspace{-5mm}
\end{center}
\end{figure*}

\noindent
\textbf{Stage 1 (S1):}
When the \DP noise is applied in S1, i.e., directly on the dataset, in Figures~\ref{fig:stageall-ACL-all}(a-b), we observe similar trends with \ACL in synthetic data.
However, \ACL remains high until $\epsilon$ increases to $\sim$$100$.
Up to that point, the modeling process is unable to learn the dataset rules, and the \ACL is indicative of random guesses from the model, and this is true regardless of the model used.
When the smallest amount of \DP noise (i.e., $\epsilon$=$1000$) is applied, we find that \NB performs the best and achieves the lowest \ACL ($0.2$$<$\ACL$<$$0.4$) between the two ML methods.
On the other hand, \NN performs the worst at this degree of \DP noise, since its \ACL is high ($0.4$$<$\ACL$<$$0.7$), regardless of dataset complexity.

\noindent
\textbf{Stage 2 (S2):}
Interestingly, as seen in Figure~\ref{fig:stageall-ACL-all}(c), when the \DP noise is applied at S2, i.e., during model training, we notice that \ACL is at its highest at $\epsilon$$\approx$$10$ for \NN.
In particular, we observe that the accuracy is generally low, and also highly dependent on the dataset used ($0.1$$<$\ACL$<$$0.9$).

\noindent
\textbf{Stage 3 (S3):}
When the \DP noise is applied at S3, i.e., after the model was trained but before it is used, we see (Figure~\ref{fig:stageall-ACL-all}(d)) the \ACL at its highest until $\epsilon$$\approx$$0.1$ for \NB.
When this inflection point is passed, and \DP is applied, the lowest \ACL$\approx$0 is achieved, and this performance is consistent across all datasets and class complexities.

\begin{remark}
\label{remark:neg_acl}
We observe that \ACL drops below $0.0$ in S3: \NB, indicating a model accuracy higher than if no privacy was applied.
It is likely that the small amounts of \DP noise applied have assisted in generalizing the model to predict better on unseen data.
However, as the \DP noise continues to increase, a diminished model performance returns.
These may be interesting cases where a practitioner can seek to obtain smaller $\epsilon$ at no cost to model performance.
\end{remark}

\noindent
\textbf{Dataset class complexity:}
Generally, we know that datasets with high class complexity are harder to model with ML methods, and thus, their accuracy achieved would be expected to be low, even in presence of no \DP noise.
Indeed, in the above experimentation, we notice that in several occasions, datasets with $50$ or $100$ classes are difficult to model with high accuracy and high \DP noise.
When small amount of \DP noise is applied on low-complexity datasets with 2, 10 or even 20 classes, and especially in S1 and S3, the tested ML methods perform fairly well, with low \ACL.

\noindent
\textbf{Comparing ML Performance Across \DP-ML} \textbf{framework}\\ \textbf{Stages:}
To offer stronger protection guarantees for the given data, more \DP noise must be added on the data (i.e., move towards the left hand-side of the aforementioned plots).
When adding more noise, it appears that the \ACL is affected in a similar fashion, for any ML method used, and regardless of the Stage at which we apply the noise, or dataset class complexity.
There is an amount of \DP noise that when it is added, it obscures much of the data variability, and consequently increases the \ACL of each trained model.
Interestingly, as identified earlier at the analysis of results from each Stage, and even on the results with synthetic data, this inflection point moves to higher levels of \DP noise (i.e., lower values of $\epsilon$), as the noise is added in later Stages in the framework.
In particular, we notice that the \ACL is drastically reduced when:
\begin{description}
	\itemsep-0.3em
	\item [Stage 1:] Inflection point of $\epsilon > 100$
	\item [Stage 2:] Inflection point of $\epsilon > 1$
	\item [Stage 3:] Inflection point of $\epsilon > 0.1$
\end{description}

Furthermore, it appears that the various models perform differently depending on the Stage the \DP noise is applied.
\NB is more effective when used at S3 than S1, for the same amount of \DP noise, the model accuracy is better (i.e., \ACL is lower).
However, if the \DP-enabled ML framework requires consistent ML performance (i.e., low \ACL) across datasets of different class complexities (i.e., 2-100 classes), then \DP noise may need to be applied at S1.
\NN performs better across all datasets when low noise is applied at S1.

\subsubsection{Membership Inference Attacks vs. \DP noise}
\label{sec:mi_results}

Next, we analyze the advantage of an attacker when mounting the \SalemMI attack, in a similar fashion as with the synthetic data, but grouping results of all real datasets by class complexity.
We discuss the effectiveness of \SalemMI attack on individual models per stage and across stages, and the impact of class complexity on the attack.

\begin{figure*}[t]
\begin{center}
\subfigure[S1: Naive Bayes]{\includegraphics[width=0.48\columnwidth]{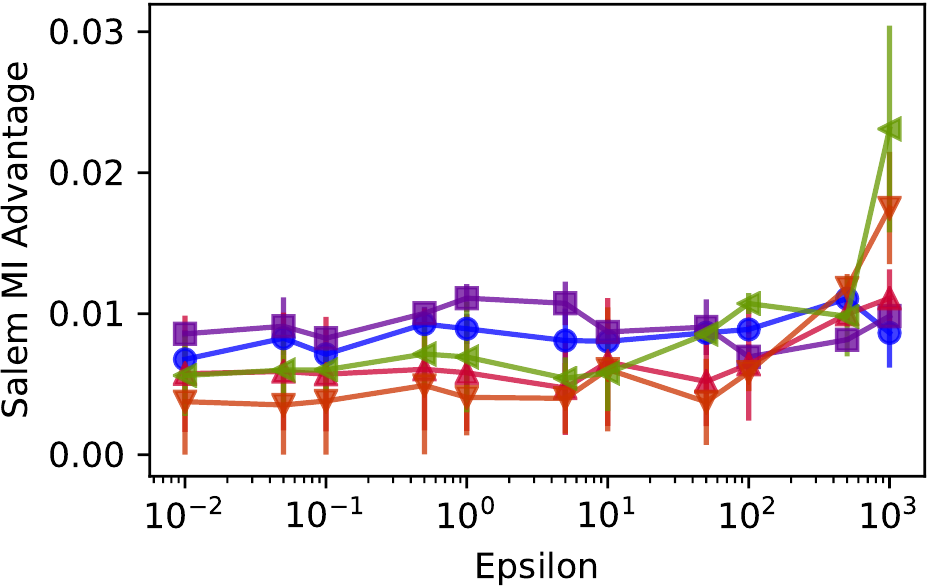}}
\hspace{2pt}
\subfigure[S1: Neural Network]{\includegraphics[width=0.48\columnwidth]{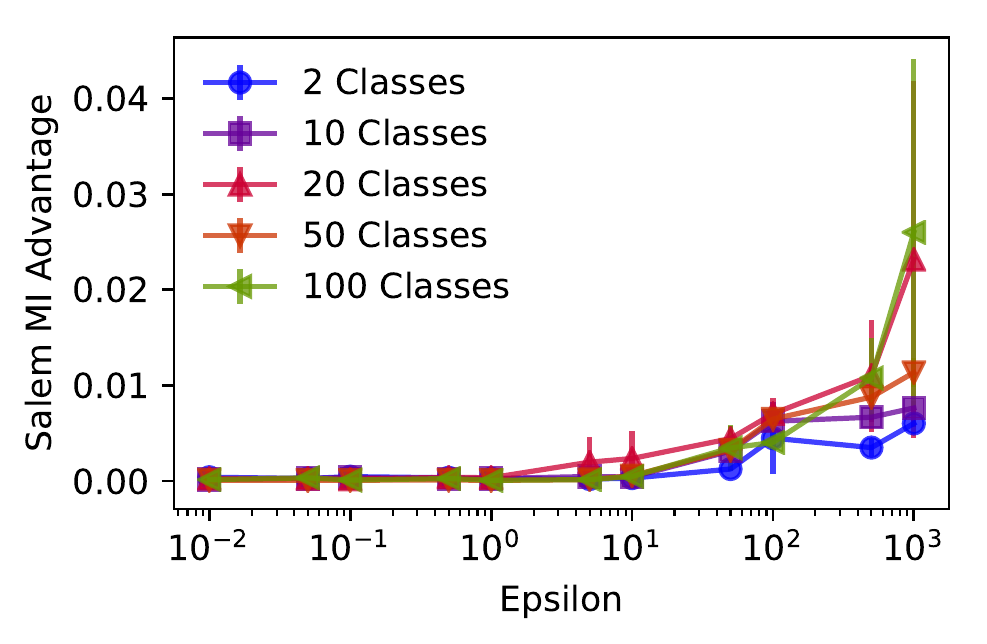}}%
\hspace{2pt}
\subfigure[S2: Neural Network]{\includegraphics[width=0.48\columnwidth]{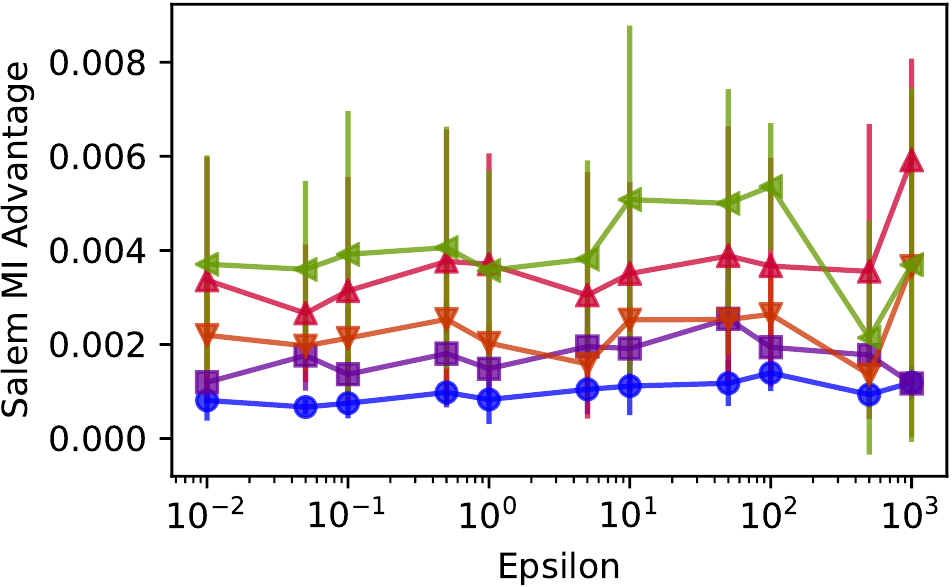}}
\hspace{2pt}
\subfigure[S3: Naive Bayes]{\includegraphics[width=0.48\columnwidth]{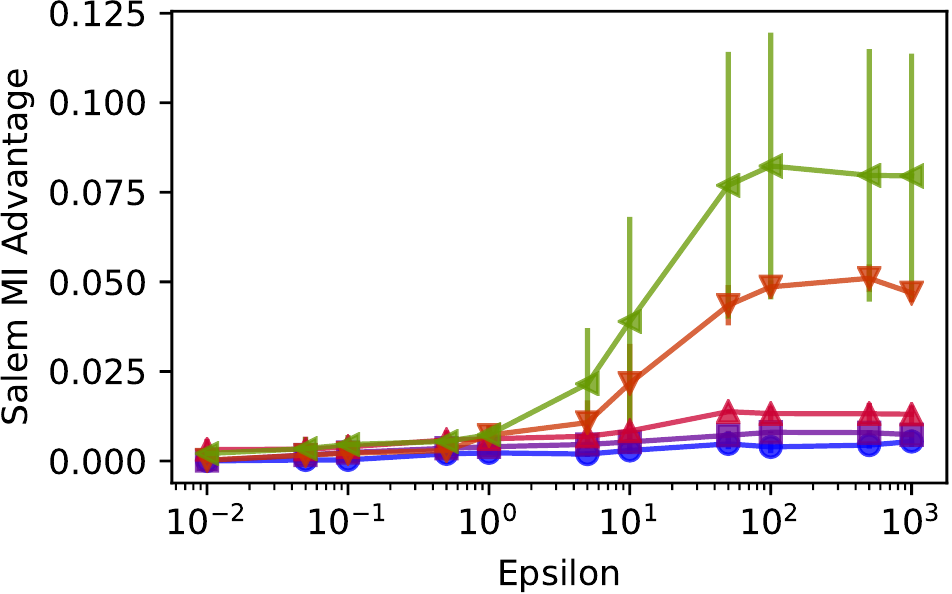}}
\vspace{-5mm}
\caption{Advantage of \SalemMI attack for each ML method used, when different amount of \DP noise is applied at Stage 1, 2 or 3 of the ML framework, and for different datasets used. We summarize the datasets by number of classes used.}
\label{fig:stageall-SalemMI-all}
\vspace{-5mm}
\end{center}
\end{figure*}

\noindent
\textbf{Stage 1 (S1):}
When \DP is applied at S1, we notice that \SalemMI advantage is generally low and close to zero, up to $\epsilon$$\approx$$100$ for \NB in Figure~\ref{fig:stageall-SalemMI-all}(a).
\NN shows a non-zero advantage from $\epsilon$$\approx$$10$ and on.
Moving from left to right in the $\epsilon$-axis, and until these thresholds are reached, the \SalemMI attacker does not gain any privacy advantage from discerning if data records were being included in the training dataset of the given model or not.

\noindent
\textbf{Stage 2 (S2):}
When \DP noise is applied at S2, the \SalemMI advantage is low for any $\epsilon$ for \NN, in Figure~\ref{fig:stageall-SalemMI-all}(c), with the effectiveness of this attack on \NN built with \DP noise added at this Stage is low.
Specifically, we observe that for the \NN, the attacker's advantage is overall low (\SalemMI $<$$0.008$), regardless of the amount of \DP noise applied.

\noindent
\textbf{Stage 3 (S3):}
When \DP is applied at S3, \SalemMI advantage increases when $\epsilon$$>$$1$ for \NB in Figure~\ref{fig:stageall-SalemMI-all}(d).
This means that when \NB is trained, datasets with high class complexity are more vulnerable. This has been previously stipulated as a result of overfitting to each specific class, given that the feature space is to be divided up into more decision regions.

Interestingly, all aforementioned results demonstrate similar patterns with the results on synthetic data (i.e., Fig.~\ref{fig:stageall-SalemMI-synth} and~\ref{fig:stageall-SalemMI-all}).

\noindent
\textbf{Comparing \SalemMI Across \DP-ML framework Stages:}
As expected, when adding less \DP noise in the framework (depending on the Stage at which it is applied), this impacts the effectiveness of a \SalemMI attacker.
In particular, when the inflection points below are reached, the attacker has a non-zero advantage.

\begin{description}
	\itemsep-0.3em
	\item [Stage 1:] Inflection point of $\epsilon$$>$$10$$\sim$$500$
	\item [Stage 2:] Inflection point of $\epsilon$$>$$0.5$
	\item [Stage 3:] Inflection point of $\epsilon$$>$$1$
\end{description}

Additionally, for the same amount of \DP noise, different ML methods allow the attacker to learn different amounts of private information (i.e., which instances of data are members of the training set).
For example, \NB allows the attacker to learn up to 10x more when the \DP is applied in S3 than in S1, in addition to the inflection point to be found in lower $\epsilon$ values for S3 than for S1.
Finally, when \DP is applied in S2:\NN, there is 10x less privacy leakage than S1:\NN.

\subsubsection{Attribute Inference Attacks vs. \DP noise}\label{sec:ai_results}

Next, we analyze the results of \YeomAI attack on real data, in a similar fashion as with the synthetic data, but again, grouping results of all real datasets by class.
Again, we discuss the attack's effectiveness on each stage and across stages, and how class complexity is an influencing factor.

\begin{figure*}[t]
\begin{center}
\subfigure[S1: Naive Bayes]{\includegraphics[width=0.48\columnwidth]{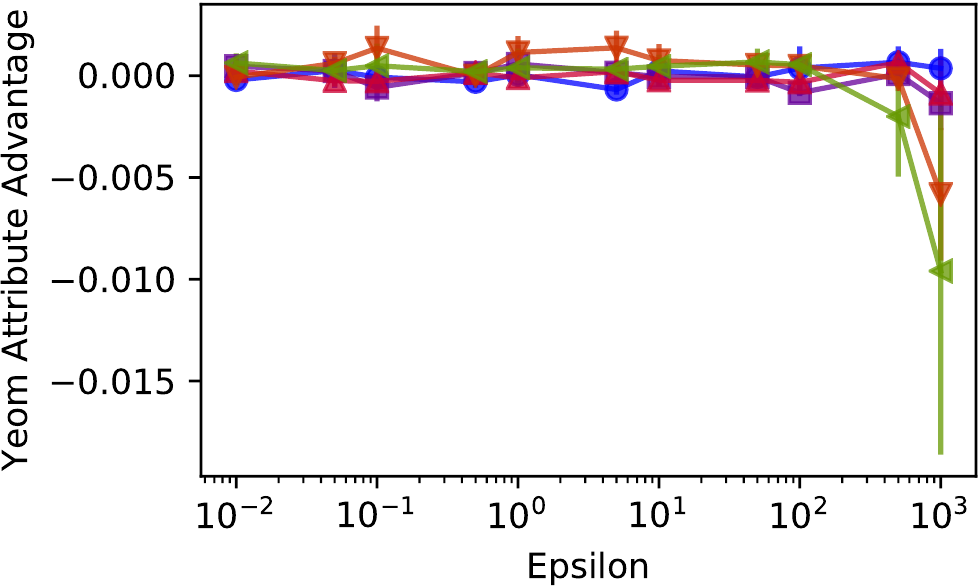}}
\hspace{2pt}
\subfigure[S1: Neural Network]{\includegraphics[width=0.48\columnwidth]{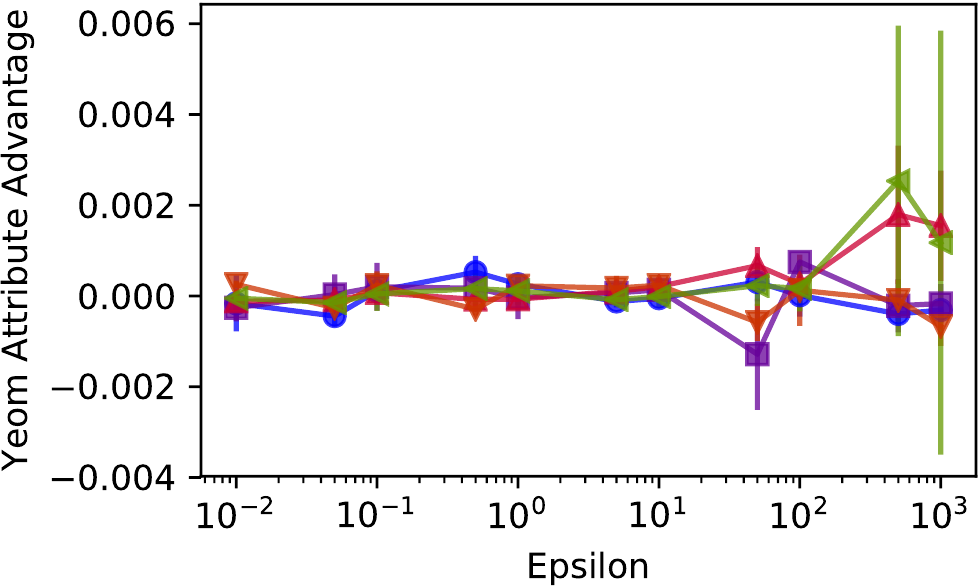}}%
\hspace{2pt}
\subfigure[S2: Neural Network]{\includegraphics[width=0.48\columnwidth]{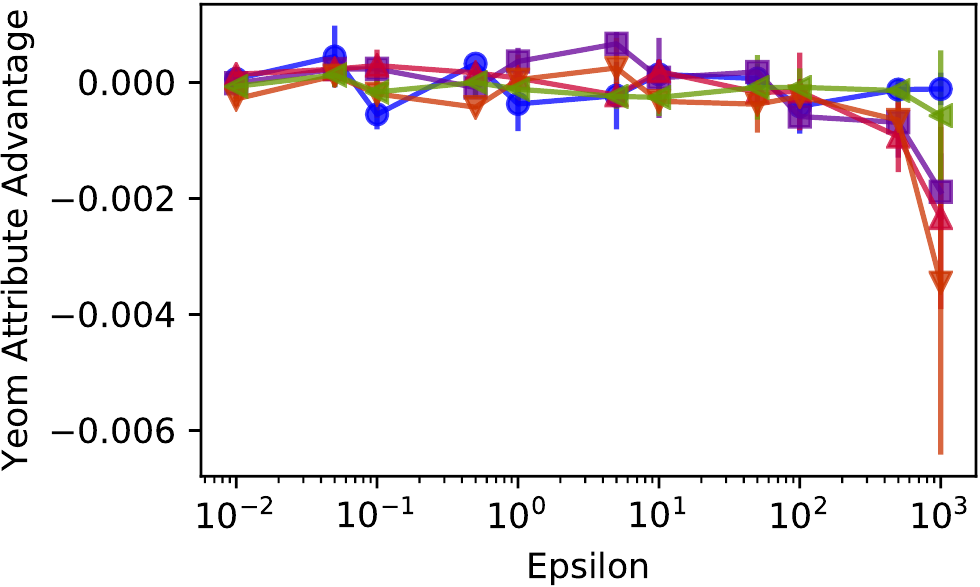}}
\hspace{2pt}
\subfigure[S3: Naive Bayes]{\includegraphics[width=0.48\columnwidth]{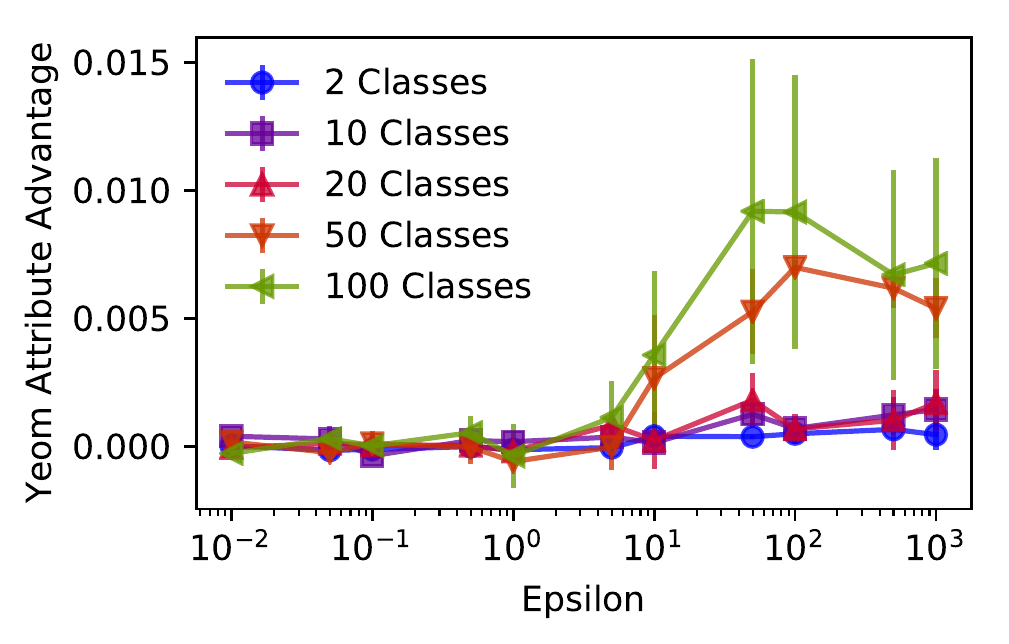}}
\vspace{-5mm}
\caption{Advantage of \YeomAI attack for each ML method used, when different amount of \DP noise is applied at S1, S2, and S3 of the framework, and for different datasets used. We summarize the datasets by number of classes used.}
\label{fig:stageall-YeomAI-all}
\vspace{-2mm}
\end{center}
\end{figure*}

\noindent
\textbf{Stage 1 (S1):}
From Figures~\ref{fig:stageall-YeomAI-all}(a-b), we observe that for many of the models, and for the different datasets and class complexities, the \YeomAI advantage is very low and even negative, which points to failed attack for leaking private information on the attributes of member data vectors in comparison to non-members.
A negative advantage indicates that the attacker can achieve greater attack success when a vector has been excluded from the training dataset than when kept within.
Therefore, this attack is not very effective when executed on a \DP-enabled ML framework that has trained models while injecting \DP noise at S1, i.e., before any ML training.
Between of the two models trained, \NN would potentially leak the most, when the \DP noise is low ($\epsilon$$>$$100$).
Interestingly, the adversary's advantage would still be 6x lower than in \SalemMI, though AI is the more difficult attack.

\noindent
\textbf{Stage 2 (S2):}
When \DP noise is added at S2, from Figure~\ref{fig:stageall-YeomAI-all}(c), we observe an equally low advantage, however the attacker achieves negative advantage (Negative advantage may not necessarily be disadvantageous, as there is still sufficient information to distinguish between members and non-members.).

\noindent
\textbf{Stage 3 (S3):}
Similarly with S2, in S3, a \YeomAI attacker can leak information about attributes of the data, when \NB is trained, and above an inflection point of $\epsilon$$>$$5$ (Figure~\ref{fig:stageall-YeomAI-all}(d)).
This advantage is 8x lower than \SalemMI.

\noindent
\textbf{Comparing \YeomAI Across \DP-ML framework Stages:}
Examining this attack across Stages, for different ML methods, we observe the following.
\NB allows the attacker a superior ability to leak more private information when \DP noise is applied in S3.
In fact, if \DP is applied at S3, the attacker will make 3x more errors (false positives) than S1, while trying to infer values of attributes.
Finally, \NN is more robust against AI attacks when \DP noise is applied at S2, in comparison to S1 which allows some information on attributes to leak at low amounts of \DP noise.
Again, we notice a clear shift of the inflection point to lower levels of $\epsilon$ as previously seen in the MI attacks:
\begin{description}
	\itemsep-0.3em
	\item [Stage 1:] Inflection point of $\epsilon$$>$$100$
	\item [Stage 2:] Inflection point of $\epsilon$$>$$50$
	\item [Stage 3:] Inflection point of $\epsilon$$>$$5$
\end{description}

\noindent
\textbf{Comparing MI and AI attacks:}
In general, we observe that AI attacks are less successful in leaking information about the data than MI attacks.
This is based on the advantages computed in the \SalemMI attack that are mostly positive and of higher values than the \YeomAI attack values achieved, which were mostly zero or negative.
This is to be expected, since an AI attack is an objectively more demanding attack with more potential for producing an incorrect result with the need to predict the exact value, instead of a binary membership/non-membership decision.
It is also more difficult to be carried out in practice, due to the prerequisite knowledge the attacker should have of all but 1 attribute values.

\subsubsection{DP-based ML under Constrained \ACL or $\epsilon$}

An ML practitioner may wish to apply the most effective ML approach while considering constraints for either the \ACL or $\epsilon$.

\noindent
\textbf{\ACL-bounded recommendations:}
We now determine which \DP-based ML algorithm offers the best privacy guarantees $(\epsilon)$, when a practitioner's accuracy requirements are constrained.
Specifically, we consider when the \ACL cannot exceed a pre-determined threshold.
To find the corresponding privacy offered ($\epsilon$) and the associated ML technique, we linearly interpolate the empirical trend of \ACL and $\epsilon$.
Then, we find the value of $\epsilon$ closest to the bounded \ACL, for all ML methods tested.
Finally, we report the lowest $\epsilon$, and the corresponding ML method.

We display results for \ACL constraints in Table~\ref{tab:constrained_acl}.
We observe that \NB with \DP noise applied in S3 is a prevalent option that can offer good accuracy for datasets with various class complexities.
Only \NN in S1 is a viable option for a binary class dataset, when the \ACL requirement is very low (e.g., 0.01).

\noindent
\textbf{$\epsilon$-bounded recommendations:}
We now determine which \DP-based ML algorithm offers the least accuracy loss, when a practitioner's privacy guarantee has been mandated.
We use a similar interpolation technique.
The results in Table~\ref{tab:constrained_eps} show that \NN with \DP noise applied in S1 are better options when high privacy constraints are required.

However, they lead to high \ACL, which renders the models useless.
When the privacy requirement can be relaxed, and the noise is applied in S2 or S3, then \NB is a better option for maintaining ML accuracy, this remains true for datasets with low or high class complexity.

\begin{table}[t]
\caption{Given a constrained \ACL, we show best attainable privacy guarantee ($\epsilon$), and the responsible \DP-ML algorithm.}
\label{tab:constrained_acl}
\vspace{-4mm}
\resizebox{1\columnwidth}{!}{%
\begin{tabular}{l|ll|ll|ll|ll|ll}
\hline
 & \multicolumn{2}{c|}{2 Classes} &  \multicolumn{2}{c|}{10 Classes} &  \multicolumn{2}{c|}{20 Classes} &  \multicolumn{2}{c|}{50 Classes} &  \multicolumn{2}{c}{100 Classes} \\
\hline
ACL & $\epsilon$ & DP-ML & $\epsilon$ & DP-ML & $\epsilon$ & DP-ML & $\epsilon$ & DP-ML & $\epsilon$ & DP-ML \\ \hline \hline
0.01 & 50.00 & S1-\NN & 16.52 & S3-\NB & 38.11 & S3-\NB & 31.71 & S3-\NB & 30.17 & S3-\NB \\
0.02 & 47.23 & S3-\NB & 14.18 & S3-\NB & 35.99 & S3-\NB & 30.01 & S3-\NB & 28.61 & S3-\NB \\
0.05 & 37.62 & S3-\NB &  9.47 & S3-\NB & 29.61 & S3-\NB & 24.89 & S3-\NB & 23.92 & S3-\NB \\
0.10 & 21.61 & S3-\NB &  7.31 & S3-\NB & 18.99 & S3-\NB & 16.37 & S3-\NB & 16.10 & S3-\NB \\
0.20 &  7.70 & S3-\NB &  4.52 & S3-\NB &  8.16 & S3-\NB &  8.32 & S3-\NB &  8.52 & S3-\NB \\
0.30 &  1.14 & S3-\NB &  3.48 & S3-\NB &  4.99 & S3-\NB &  5.64 & S3-\NB &  6.08 & S3-\NB \\ \hline
\end{tabular}
}
\vspace{-1mm}
\end{table}

\begin{table}[t]
\caption{Given a constrained $\epsilon$, we show the smallest compromise in \ACL, and the responsible \DP-ML algorithm.}
\label{tab:constrained_eps}
\vspace{-4mm}
\resizebox{1\columnwidth}{!}{%
\begin{tabular}{l|ll|ll|ll|ll|ll}
\hline
 & \multicolumn{2}{c|}{2 Classes} &  \multicolumn{2}{c|}{10 Classes} &  \multicolumn{2}{c|}{20 Classes} &  \multicolumn{2}{c|}{50 Classes} &  \multicolumn{2}{c}{100 Classes} \\
\hline

$\epsilon$ & ACL & DP-ML & ACL & DP-ML & ACL & DP-ML & ACL & DP-ML & ACL & DP-ML \\ \hline \hline
0.01 & 0.321 & S1-\NN & 0.804 & S1-\NN & 0.863 & S1-\NN & 0.950 & S1-\NN & 0.958 & S1-\NN \\
0.10 & 0.321 & S1-\NN & 0.802 & S1-\NN & 0.858 & S3-\NB & 0.949 & S1-\NN & 0.952 & S3-\NB \\
1.0 & 0.301 & S2-\NB & 0.540 & S3-\NB & 0.634 & S3-\NB & 0.717 & S3-\NB & 0.727 & S3-\NB \\
10 & 0.136 & S3-\NB & 0.038 & S3-\NB & 0.142 & S3-\NB & 0.137 & S3-\NB & 0.139 & S3-\NB \\
100 & 0.001 & S3-\NB & -0.141 & S3-\NB & -0.055 & S3-\NB & -0.124 & S3-\NB & -0.135 & S3-\NB \\
1000 & -0.001 & S3-\NB & -0.127 & S3-\NB & -0.042 & S3-\NB & -0.109 & S3-\NB & -0.121 & S3-\NB \\ \hline

\end{tabular}
}
\vspace{-1mm}
\end{table}

\subsubsection{Summary of Findings}
\label{sec:summary_of_findings}

\begin{figure*}[t]
    \centering
    \subfigure[\SalemMI]{\includegraphics[width=1\columnwidth]{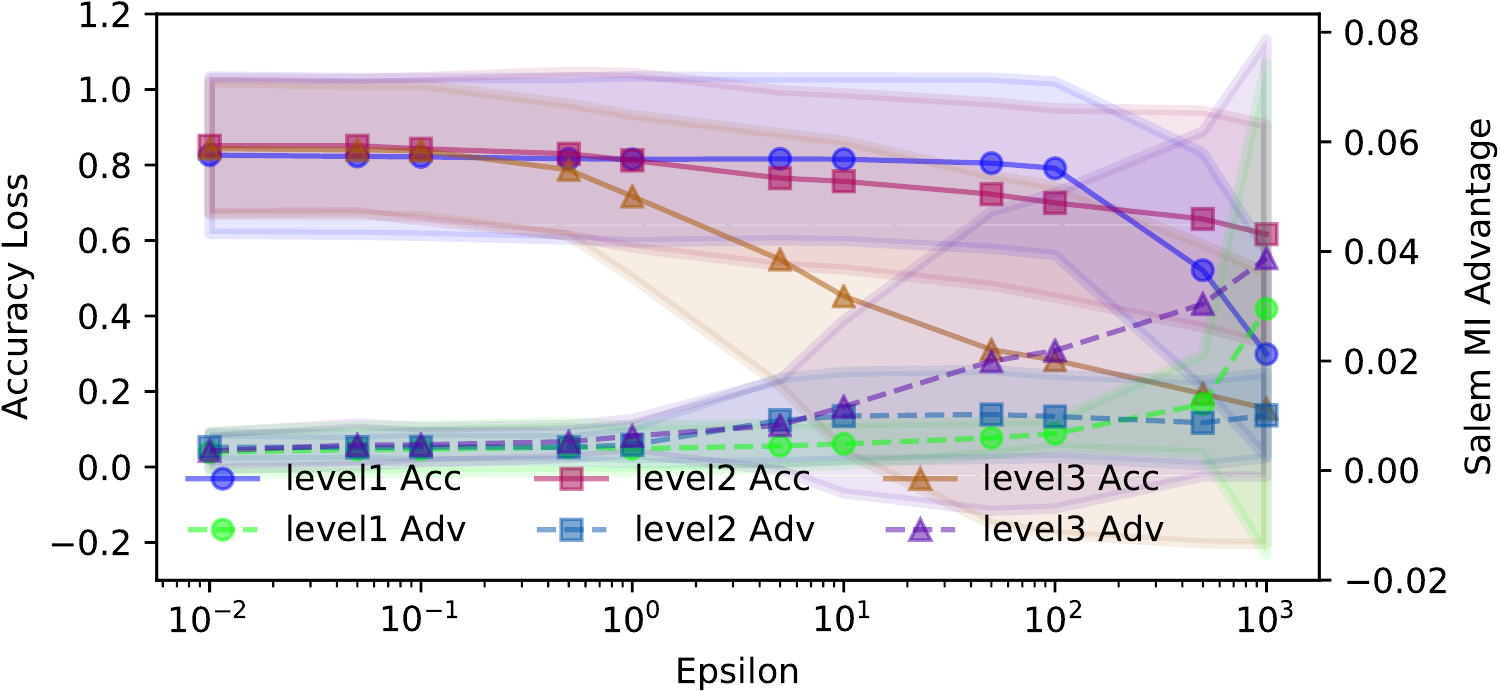}} 
    \hspace{2mm}
    \subfigure[\YeomAI]{\includegraphics[width=1\columnwidth]{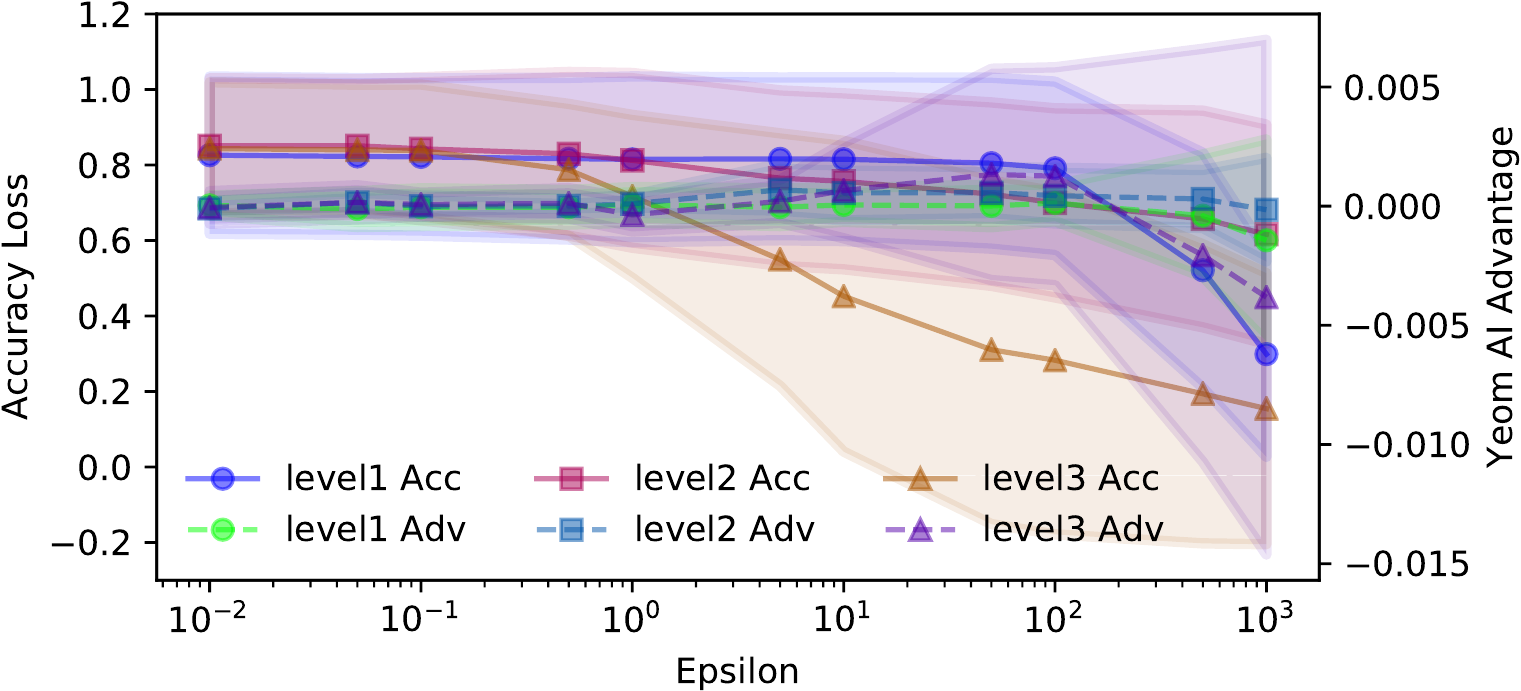}}
	\vspace{-4mm}
	\caption{Summary plot of \ACL (y1-axis) and Privacy advantage (y2-axis) vs. $\epsilon$ applied (x-axis), for each Stage.
	Each point, for a line of a given Stage, is the mean across all results for different ML methods and datasets.
	Shaded colored areas signify 1 st. dev. around each mean.
	}
    \label{fig:all-stages-ACL-SalemMI-YeomAI}
\end{figure*}

In Figure~\ref{fig:all-stages-ACL-SalemMI-YeomAI}, we summarize the findings from different experimental setups, for \ACL and for the two privacy attacks of \SalemMI and \YeomAI.
In these summary figures, the tradeoff between \ACL and protection against privacy leaks emerges more clearly.
From this figure, and based on all previous explorations with respect to \ACL and the two privacy attacks (\SalemMI, and \YeomAI), we summarize our key takeaways:
\vspace{-1mm}
\begin{enumerate}
\itemsep0em

\item For a given amount of \DP noise applied, ML models predict better (i.e., have good accuracy and low \ACL), when the noise is inserted at a later Stage (e.g., S2 or S3 than S1) [Sec.~\ref{sec:acl-results}].
\item Conversely, to achieve reduced privacy leaks (in lower attack advantages) with least amount of \DP noise, this noise must be added at earlier Stages in the framework (S1 > S2 > S3) [Sec.~\ref{sec:mi_results} \& \ref{sec:ai_results}]. Additionally:
{
\setlength\itemindent{25pt}\setlength\itemsep{0pt}\indent
\indent \item The amount of \DP noise added to a given DP-ML method does not influence the inflection point of a privacy attack (both MI and AI); instead, the inflection of attack success is dependent on the DP-ML method used and framework Stage the noise is applied (as noted in Takeaway 2).
}
\item In both synthetic and real datasets, the data complexity is demonstrably unlikely to affect the inflection point of \ACL [Sec.~\ref{sec:acl-results}], or attack advantage [Sec.~\ref{sec:mi_results} and~\ref{sec:ai_results}] for a given \DP-ML method. We also  corroborate the known result that higher class complexities (more classes) produce higher privacy leaks~\cite{shokri2017membership,salem2018ml}.

{
\setlength\itemindent{-12pt}
\indent In the context of both \ACL and Privacy:
}
{
\item The performance of current state-of-art MI and AI attacks is directly related to the prediction accuracy of the DP-ML model used.
The inflection points of \ACL and privacy advantage for each DP-ML method correspond to approximately the same amount of \DP noise.
}
\item When investigating the tradeoff over a wide range of \ACL and $\epsilon$ constraints, we observe that S3:\NB is the superior performing \DP-ML method.
\item Evaluating the privacy-utility tradeoff with synthetic [Sec.~\ref{sec:synth-result}] and real-world data [Sec.~\ref{sec:real_data}] yields similarities in trends and takeaways.
There is potential for a dataset-agnostic approach to estimate inflection points for similarly classed data.
\end{enumerate}

\section{DISCUSSION}\label{sec:discussion}

We presented a comprehensive empirical study on the inherent tradeoff between utility and privacy when applying \DP on ML algorithms.
We investigated two, state of art \DP-enabled ML and DL algorithms currently available in literature, and evaluated the aforementioned tradeoff in each ML method, using two privacy inference attacks and one utility metric.
We performed this investigation using both synthetic datasets and three commonly used real datasets of varying class and attribute complexity.
Finally, we extracted from this experimentation various lessons, and offered recommendations to interested privacy ML researchers.

During this evaluation with our framework, we limited the number of experimental configurations, to make the problem tractable with comparable results.
Next, we discuss experimental variants that can be investigated in the future with our framework.

\textbf{\DP variants:}
In this study, we considered only $\epsilon$-\DP.
However, as already mentioned in Section~\ref{sec:dp-definitions}, there is an increasing number of \DP compositions and relaxations, such as $(\epsilon, \delta)$-\DP and $(\alpha, \epsilon)$-\DP.
Interestingly, these \DP relaxations are relatively recent, and many of the \DP-enabled ML algorithms available in literature that we used, are still using the original $\epsilon$-\DP.
Future work should address how to adapt such algorithms to support newer \DP relaxations, but should also enable ML practitioners to fairly compare these methods.
For this, one would need to establish an equivalence between the various \DP options available.
In fact, in the future, even a simple evaluation of how a varying $\delta$ in $(\epsilon, \delta)$-\DP impacts the resulting \ACL and privacy metrics would be highly informative.

\textbf{Local vs. Global \DP:}
The boundary of trusted and non-trusted entities is becoming increasingly blurred.
On one hand, ML model holders seek to protect their models' privacy and user data.
On the other, privacy advocates argue even the model holders should not be a trusted entity.
In fact, there is a notion of trust in the \DP ML pipeline:
\textit{Local \DP} is when \DP is applied very close to data generation without considering information or context about the entire system. 
Instead, \textit{Global \DP} does not need to tradeoff as much utility for same mathematical guarantees: with global system view, it can make more intelligent decisions on how to apply \DP noise.
In our framework, Local \DP loosely corresponds to inserting \DP noise in S1, with ML training receiving \DP-protected data, whereas Global \DP corresponds to \DP noise applied in S2 or S3, with the model having unfettered access to unprotected data.
\textbf{Utility metrics:}
We focused on accuracy (loss) of a \DP-enabled ML model with respect to its non-private counterpart.
However, as mentioned in Section~\ref{sec:utility-metrics}, more metrics can be employed to assess the change in utility of a trained \DP-enabled ML model, such as precision, recall, F1 measure, etc.
Furthermore, model Fairness~\cite{corbett2018measure} is another metric of particular interest given the increased public scrutiny of ML model fairness in the context of well established anti-discriminatory legal frameworks across the globe.

\textbf{Computation Cost:}
Another potential aspect of the privacy-utility tradeoff to be studied is the resource overhead and its relation to the amount of \DP noise, and the framework Stage it is added.

\section{Future Work}
In this paper, we have laid the groundwork for systematically evaluating Differentially-Private (DP) machine learning techniques.
The next steps will be to include in our study additional learning methods such as DP-Logistic Regression~\cite{nb-lr-dp, zhang2012functional} and DP-Random Forests~\cite{fletcher2017differentially}. In additional to new algorithms, newly proposed approaches to existing methods, for example adaptive weight clipping in stochastic gradient decent (\NN)~\cite{pichapati2019adaclip}, can improve utility whilst preserving the DP guarantees.
Additionally, in the current landscape of Inference attacks, researchers continue to increase the effectiveness of said attacks, with alternate MI attacks, like the one of Shokri's~\cite{shokri2017membership} or Yeom's MI attack \cite{yeom2018privacy}, or new AI attacks~\cite{zhao2019inferring}.
From our current investigation, we were able to demonstrate a clear inflection point of the tradeoff between \textit{Utility} and \textit{Privacy}, in relation to the amount of privacy added (with DP-noise).
Of particular interest, is the ability to predict this behavior either through an expanded set of generic benchmarks, or through modeling the relationship between the data, privacy, and utility of the ML pipeline.

\section{CONCLUSION}\label{sec:conclusion}

Privacy-preserving machine learning (ML) methods come with the inherent tradeoff between model utility achieved and privacy offered by the technique applied to protect the data.
Our main contribution with this paper is the proposal of a practical evaluation framework that enables a privacy ML researcher to study this tradeoff in depth for their data, and make data-driven decisions on where to apply Differentially Private (\DP) noise inside their ML framework to protect their data and model, while achieving the best possible ML accuracy.

We identify three such Stages in the \DP-enabled ML framework where \DP-based noise can be added:
1) directly at the data collection,
2) during model training, or
3) at model finalization.
We allow the practitioner to apply different amounts of noise based on the privacy guarantees they have, and at the different stages in the framework, and study the aforementioned tradeoff between utility of the model trained, and the privacy of the data or model achieved, using two well-known privacy attacks.

We use our framework to comprehensively evaluate various implementations of \DP-based ML algorithms, and measure their ability to fend off real-world privacy attacks, in addition to measuring their core goal of providing accurate classifications.
We evaluate each implementation across a range of privacy budgets, and datasets, each implementation providing the same mathematical privacy guarantees.
By measuring the models' resistance to real world attacks of membership and attribute inference, and their classification accuracy, we determine which methods provide the most desirable tradeoff between privacy and utility.
Building on our results, we provide recommendations to a privacy ML researcher on how to select appropriate, \DP-based ML methods, based on the data complexity at hand, and privacy guarantees and utility needs.

\section*{Acknowledgments}
The research leading to these results has received partial funding from Optus Macquarie University Cybersecurity Hub, Data61 CSIRO and an Australian Government Research Training Program (RTP) Scholarship.
It has also received funding from the EU’s Horizon 2020 Programme under grant agreements No 830927 (project CONCORDIA), No 871370 (project PIMCITY) and No 871793 (project ACCORDION).
The paper reflects only the authors’ views and the Agency and the Commission are not responsible for any use that may be made of the information it contains.
The authors would like to thank the anonymous reviewers for their feedback to improve on the ongoing research.
\bibliographystyle{ACM-Reference-Format}
\bibliography{ref.bib}

\end{document}